\documentclass[fleqn,usenatbib]{mnras}

\usepackage{newtxtext,newtxmath}

\usepackage[T1]{fontenc}
\usepackage{ae,aecompl}

\usepackage[mathscr]{euscript}
\let\euscr\mathscr \let\mathscr\relax
\usepackage[scr]{rsfso}

\DeclareMathAlphabet{\mathpzc}{OT1}{pzc}{m}{it}

\usepackage{graphicx}	
\usepackage{amsmath}	
\usepackage{amssymb}	

\title[A search for cosmological anisotropy using SDSS]{A search for cosmological anisotropy using the Lyman alpha forest from SDSS quasar spectra}

\author[E.\,O. Zavarygin and J.\,K. Webb]{
Evgeny O. Zavarygin\thanks{E-mail: e.zavarygin@gmail.com (EOZ)}
and John. K. Webb\thanks{E-mail: jkw@phys.unsw.edu.au (JKW)}
\\
School of Physics, University of New South Wales, Sydney, NSW 2052, Australia
}

\date{Accepted XXX. Received YYY; in original form ZZZ}

\pubyear{2019}

\begin{document}
\label{firstpage}
\pagerange{\pageref{firstpage}--\pageref{lastpage}}
\maketitle

\begin{abstract}
The Cosmological Principle, the combined assumptions of cosmological isotropy and homogeneity, underpins the standard model of Big Bang cosmology with which we interpret astronomical observations. A new test of isotropy over the redshift range $2<z<4$ and across large angular scales on the sky is presented. We use the cosmological distribution of neutral hydrogen, as probed by the Ly\,$\alpha$ forest seen towards distant quasars. The Sloan Digital Sky Survey provides the largest dataset of quasar spectra available to date. We use combined information from Data Releases 12 and 14 to select a sample of 142,661 quasars most suitable for this purpose. The scales covered by the data extend beyond post-inflation causality scales, thus probing initial conditions in the early universe. We identify significant spatially correlated systematic effects that can emulate cosmological anisotropy. Once these systematics have been accounted for, the data are found to be consistent with isotropy, providing an important independent check on the standard model, consistent with results from cosmic microwave background data.
\end{abstract}




\section{Introduction}

The extent to which the Cosmological Principle remains valid as one goes to increasingly large scales in the universe is an important question given several important anomalies that claim to detect deviations from a perfect Cosmological Principle \citep{Schwarz2016}. 

The recent dramatic improvements in astronomical instrumentation and facilities allow increasingly stringent tests.  The large-scale distribution of hydrogen clouds, the Ly\,$\alpha$ forest, seen in absorption against background high redshift quasars, provides a unique cosmological database and permits a new test of isotropy. In analogy to the cosmoc microwave background (CMB) horizon problem, one can calculate the comoving particle horizon at the observed redshift for comparison with the angular size of coherent features in the Ly\,$\alpha$ forest as seen in the plane of the sky from Earth. The idea of using Ly\,$\alpha$ forest as a fundamental test of this sort was introduced by \cite{Rauch1994}. Here we modify the original idea, adapting it to suit the Sloan Digital Sky Survey \citep[SDSS;][]{Gunn2006} quasar survey. 

The comoving particle horizon for a ``fundamental observer'' who exists at some cosmic instant $t$ defines a distance beyond which no causal effects could ever have existed \citep{Rindler1956}, i.e. since inflation and assuming a constant speed of light $c$. This is given by
\begin{equation}
\euscr{P}_h = c \int\limits_0^t \frac{dt}{a(t)} = c \int\limits^{\infty}_{z_{\rm obs}} \frac{dz}{H(z)}
\label{eq:particlehorizon}
\end{equation}
where $a$ is the scale factor, $z_{\rm obs}$ is the redshift of the imagined fundamental observer as measured from Earth,\\ $H(z)=H_0\sqrt{\Omega_{\textsc{r}}(1+z)^4 + \Omega_{\textsc{m}}(1+z)^3 + \Omega_{\Lambda}}$ is the Hubble parameter at redshift $z$ which takes the value $H_0$ at the present epoch, and $\Omega_{\textsc{r}}$, $\Omega_{\textsc{m}}$, and $\Omega_{\Lambda}$ are the CMB radiation, non-relativistic matter and the dark energy densities of the Universe at the present epoch.

\begin{figure*}
\centering
\includegraphics[width=0.9\linewidth]{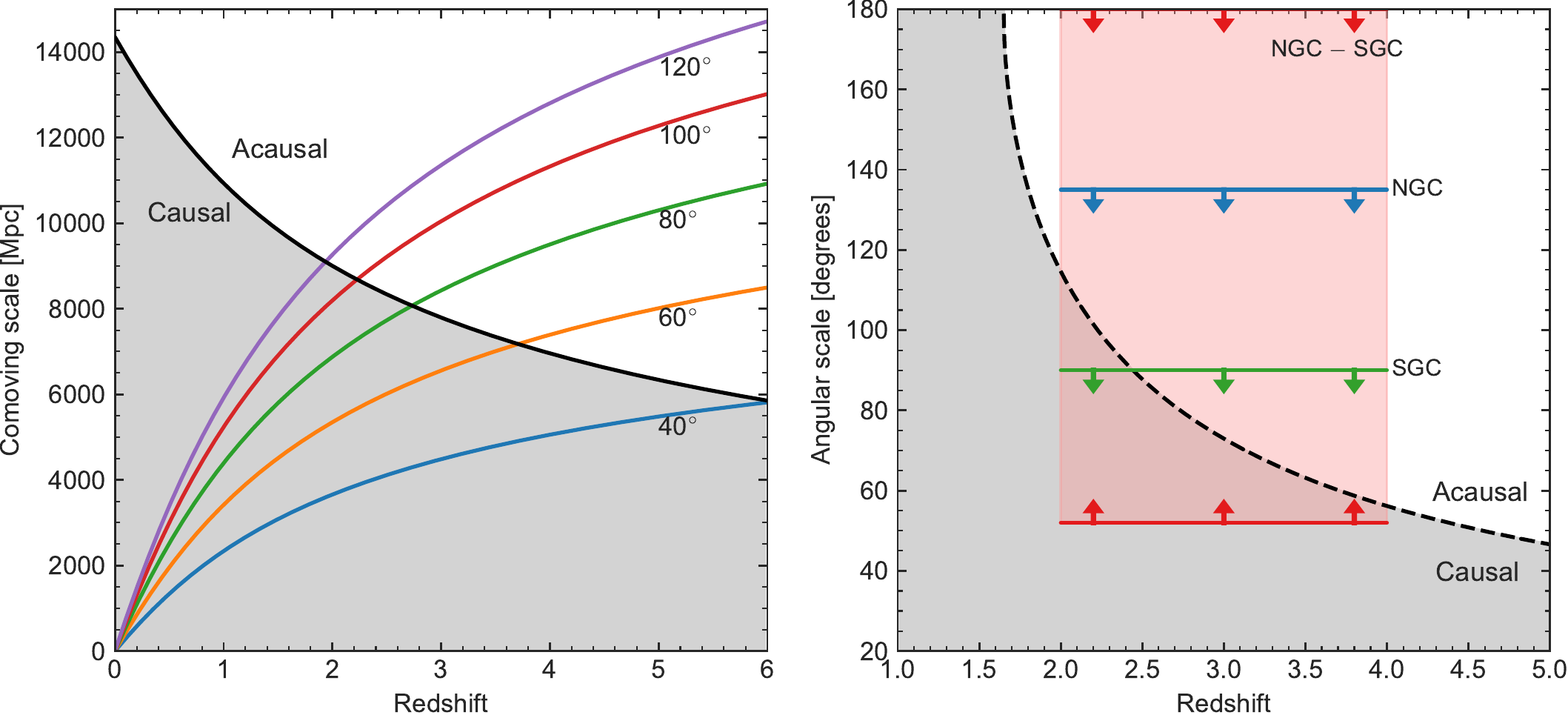}
\caption{{\it Left panel:} The coloured lines emanating from coordinate (0,0), given by equation~(\ref{eq:angularhorizon}), show the comoving separation between two points on the plane of the sky subtending an angle $\theta$, as viewed by an observer at the present epoch. Curves are plotted for five different angles on the sky. The black line shows the particle horizon, equation~(\ref{eq:particlehorizon}). Cosmological parameters by \protect\cite{Planck2015XIII}  are used. At redshifts higher than the intersection points between the black and coloured lines, regions separated by the angle corresponding to the coloured line are acausal i.e. if the speed of light is constant and if inflation did not take place, we would have no reason to expect mean Ly\,$\alpha$ forest properties to match. {\it Right panel:} Angular scale of the causal boundary.  The dashed black curve corresponds to equality between comoving separation (coloured lines in the left panel) and the comoving particle horizon (black line in the left panel) i.e. the dashed black line shows the intersection described above. The translucent red shaded area corresponds to the range in angular separations for quasar pairs from two patches in the sky, NGC and SGC (Section~\ref{seC:data_release}). For the NGC-SGC comparison discussed in Section~\ref{sec:ngc_sgc}, the data are predominantly acausal. The horizontal lines labelled NGC and SGC indicate the maximum angular extent of each patch.}
\label{fig:sdss_causality_scales}
\end{figure*}

The comoving separation of two points at the same redshift $z$ on the plane of the sky subtending an angle $\theta$, as viewed by an observer at the present epoch, is
\begin{equation}
\euscr{L}_{\theta} = 2 c \sin \left( \theta/2\right) \int\limits^{z_{\rm obs}}_0 \frac{dz}{H(z)}.
\label{eq:angularhorizon}
\end{equation}
Fig.~\ref{fig:sdss_causality_scales} (left panel) illustrates equations~(\ref{eq:particlehorizon}) and (\ref{eq:angularhorizon}). 
At the lower end of the redshift range of typical ground-based Ly\,$\alpha$ forest observations (i.e. $z \sim 2$), acausality exists only for angular separations on the sky larger than about 100 degrees. By $z=4$ that angle has dropped to about 55 degrees. Fig.~\ref{fig:sdss_causality_scales} (right panel) illustrates the angular scale at which the particle and angular horizons match (i.e. equality of equations~\ref{eq:particlehorizon} and \ref{eq:angularhorizon}). By comparing the statistical characteristics of H\,{\sc i} clouds over angular scales greater than these values, we therefore obtain information about the initial conditions in the universe. 

\section{Astronomical data}

\subsection{Survey, data release and quasar catalogue}
\label{seC:data_release}

The quasar spectra used are from the SDSS. We use the SDSS Data Release 12 (DR12) quasar catalogue \citep[DR12Q;][]{Paris2017}, providing a list of 297,301 quasars covering 9,376 square degrees, corresponding to 23\% of the sky.  The Ly\,$\alpha$ forest redshift range covered by these data is approximately $2 < z < 6$, corresponding to 0.9 to 3.3 Gyr after the Big Bang for a standard $\Lambda$CDM cosmology. Each sightline intersects hundreds of Ly\,$\alpha$ clouds.  The 3-dimensional volume probed by the data used in the analysis described here amounts to $435$ Gpc$^3$. Since the empirical (i.e. SDSS) distribution of quasar redshifts falls sharply from its peak at around $z=2.3$, if we instead use the depth range $2 < z < 4$, the volume sampled is $236$ Gpc$^3$, or 2\% of the volume included within the radial distance to recombination at $z=1100$. Such a large effective survey volume provides a cosmological probe of unprecedented sensitivity with which to test fundamental assumptions underpinning the standard cosmological model. The SDSS sky coverage is naturally divided into two distinct patches: north Galactic cap (NGC) and south Galactic cap (SGC). The angular separation between the median NGC and SGC quasar positions is approximately 144 degrees. The right-hand panel of Fig.~\ref{fig:sdss_causality_scales} illustrates that most of the areas covered by the NGC and SGC contain sight-lines along which, in the redshift range $2<z<4$, no post-inflation causal connection has ever existed. 

Subsequent to the DR12 release, the data reduction pipeline was significantly improved. Whilst damped Lyman alpha (DLA) identification was available for DR12Q, no DLA identification was available for the later quasar release, DR14Q \citep{Paris2018}, at the time we carried out this analysis.  Therefore we employed the DR12Q quasar list but made use of spectra processed with the DR14 pipeline. We used quasar emission redshifts and other metadata reported in DR14Q when available. For quasars missed in DR14Q but present in DR12Q, we use metadata from the latter.  Data quality, reduction procedures, observational details, are described extensively elsewhere \citep{Dawson2013} and we summarise only the key characteristics here: the median spectral signal-to-noise (S/N) per pixel is $\sim 3.5$, with 89\,\% of the whole sample having a S/N per pixel of $< 10$. The blue and red spectrograph arms cover the observed wavelength ranges $3560 < \lambda < 10{,}400$\,{\AA}. The spectral resolution is $R = \lambda/\Delta\lambda \sim 2000$, varying across the wavelength range covered and designed in general to give $\sim 3$ pixels per resolution element \citep{Smee2013}. Individual Ly\,$\alpha$ forest absorption lines are unresolved.

\subsection{Data pre-processing}

In each spectrum, only valid pixels are used as set by the pipeline. DLAs and sharp telluric features have been masked out as described in Sections~\ref{sec:dla_masking} and \ref{sec:sky_mask}. The previously reported flux calibration artefacts \citep{Lee2013,Bautista2017} have been corrected using a purposely-defined correction vector as described in Section~\ref{sec:corr_vector}. All the quasar spectra have been corrected for dust extinction using the dust map by \cite{Schlegel1998} and the de-reddening law from \cite{Fitzpatrick1999}. This dust correction is applied subsequent to the flux calibration correction and before the continuum fitting procedure which will be descibed in Section~\ref{sec:continuum}. To investigate the robustness of our results to dust corrections we also tried applying a different correction using the recent dust map by \cite{Planck2013XI} and our results were found insensitive to the dust corrections.

\subsubsection{Removing DLA absorption systems and associated metal lines}
\label{sec:dla_masking}

The H\,{\sc i} profiles associated with DLA absorption systems are very broad, spanning thousands of km/s.  Strong associated metal lines are seen from multiple species, some of which fall in the Ly\,$\alpha$ forest. We remove these features as follows.

Two DLA catalogues from \cite{Noterdaeme2009b,Noterdaeme2012b} and \cite{Garnett2017} provide DLA absorption redshifts and neutral hydrogen column densities. The former contains 34{,}050 DLA's distributed over 26{,}312 sightlines. The latter, instead of providing a list of DLA's, defines the probabilities for each sightline to have a DLA along with redshifts and column densities. We use these probabilities to calculate Bayes factors ($B$). With the scale of \cite{Kass1995}, we consider real only those DLA's that have a very strong evidence of empirical support, i.\,e.\ $2\ln B>10$. This leads to 35{,}848 sightlines with a DLA from the Garnett et al catalogue.

For all the DLA's from both catalogues, we mask all the pixels at wavelengths around Ly\,$\alpha$ where residual flux $I/I_{\circ} = e^{-\tau} < 95\%$. When a DLA is common to both catalogues, the resulting mask is a union of the two individual masks. The following approximation for a Voigt profile is used:
\begin{equation}
\tau = N\sigma = N \frac{e^2}{m_{\rm e}c^3}\frac{\Gamma\lambda_{\alpha}}{4\pi} f_{\alpha}\lambda_{\alpha}\left(\frac{\lambda}{\Delta\lambda}\right)^2,    
\end{equation}
where $N$ is DLA's H\,{\sc i} column density, $e$ is the electron charge (in esu units), $m_{\rm e}$ is the electron mass, $c$ is the speed of light; $\Gamma$, $\lambda_{\alpha}$, and $f_{\alpha}$ are the sum of the Einstein $A$ coefficients, the central wavelength (1215.67~{\AA}), and the oscillator strength of the Ly\,$\alpha$ transition, respectively, and $\Delta\lambda = \lambda-\lambda_{\alpha}$ is a deviation from the line center.

In addition to masking out regions containing damped Ly\,$\alpha$ absorption, we also mask regions corresponding to absorption by the most commonly seen metal transitions, where those regions fall into our Ly\,$\alpha$ forest region of interest. The following metal transitions are used: O\,{\sc i} 1302, C\,{\sc ii} 1036, 1334, C\,{\sc iv} 1548, 1550, Si\,{\sc ii} 1190, 1193, 1260, 1304, 1526, Si\,{\sc iii} 1206, Si\,{\sc iv} 1393, 1402, Al\,{\sc ii} 1670, Al\,{\sc iii} 1854, 1862, Fe\,{\sc ii} 1144. We exclude $\pm 200$~km/s centered on the DLA redshift (an average velocity width for typical DLA kinematic structure, \citealt{Prochaska1997,Zwaan2008}). 

\subsubsection{Masking telluric emission and absorption features}
\label{sec:sky_mask}

\begin{figure}
\centering
\includegraphics[width=1.0\linewidth]{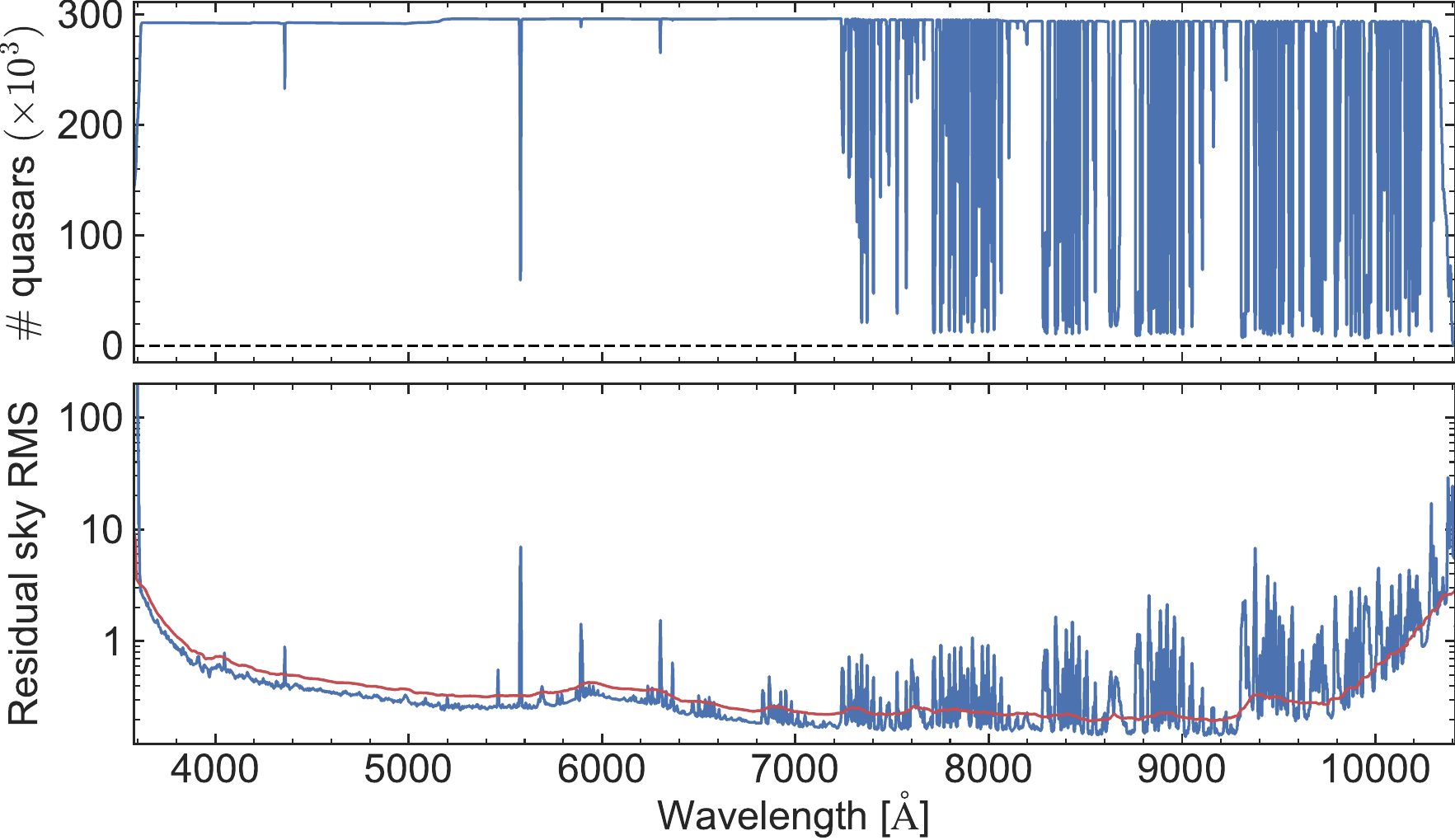}
\caption{{\it Top panel:} The number of quasars (in thousands) from the DR12Q catalogue that have a valid pixel, as set by the pipeline (non-zero inverse variance and the \texttt{and\_mask} bit set to zero), at each wavelength. {\it Bottom panel:}  The residual sky flux RMS (units: $10^{-17}$erg\,s$^{-1}$\,cm$^{-2}$\,{\AA}$^{-1}$) in the sky fibres (blue) and the accepted threshold (red). Pixels with RMS above the threshold are discarded from the analysis. Two neighbouring pixels on either side of the discarded pixels shortwards of 6500~{\AA} are also discarded. Finally, pixels outside of the wavelength range $3600-10300$~{\AA} are discarded. All pixels flagged as described above form the sky mask used in this analysis.}
\label{fig:sdss_sky_mask}
\end{figure}

Some pixels are inevitably contaminated by telluric emission/absorption. Although the data reduction pipeline process automatically flags some of the affected regions for removal (by setting zero inverse variance for contaminated pixels or/and setting the \texttt{and\_mask} bit to 23), we still found a significant number of pixels that were clearly affected and needed identifying and removing. Fig.~\ref{fig:sdss_sky_mask} (top panel) shows the number of spectra that have a valid pixel, as flagged by the pipeline, at each wavelength for all 297,301 quasars from the DR12Q catalogue.  
The Figure shows that even in very badly affected regions, some pixels still survive the automatic pipeline flagging.  If left as valid pixels, these regions cause errors to the continuum fitting process. This propagates through to the Ly\,$\alpha$ forest mean transmission estimates. To eliminate these wavelength regions affected by sky features, we use a technique similar to one described by \cite{Lee2013}. First, the RMS of the flux in the sky fibres is calculated. All 228,117 sky fibres from the DR12Q plates that have \texttt{ZWARNING} flag 1 unactivated and flag 2 activated are used. Flag 1 indicates a lack of wavelength coverage. Flag 2, when unactivated, indicates a statistically confident object detection. These cuts discard only 2\% of the sky fibres. Second, the RMS of the sky fibres is smoothed using a boxcar filter of width 31 pixels. Then, pixels with RMS greater than 1.25 times the smoothed RMS function are flagged. This is repeated iteratively until convergence. The final threshold is shown in the bottom panel of Fig.~\ref{fig:sdss_sky_mask}. Third, two pixels on either side of the flagged pixels for wavelengths below 6500~{\AA} (where we measure the Ly\,$\alpha$ forest statistics) are also flagged. Finally, pixels outside of the range $3600-10{,}300$~{\AA}, are flagged. All the pixels flagged in these four steps form the sky mask and are not used in our analysis. The overall procedure results in there being no valid pixels below 3614~{\AA} which thus defines our lower limit on quasar emission redshift.

\subsubsection{Quasar spectral shape calibration using standard star spectra}
\label{sec:corr_vector}

The pipeline reduction process includes the removal of instrumental sensitivity effects and correcting the observed quasar spectra such that intrinsic spectral shapes are recovered.  This is done using correction functions derived from standard star spectra.  Refinements to this pipeline process are discussed in section 5.2 of \cite{Lee2013}. The pipeline procedure provides a PCA model for each spectrum, used for object classification (galaxy, quasar or star) and redshift measurements \citep{Bolton2012}. The ratio of the flux values in the actual spectrum to those in the PCA model is expected to be 1 on average. However, systematic deviations are seen, and the following procedure is used to measure a correction vector.

\begin{figure}
\centering
\includegraphics[width=1.0\linewidth]{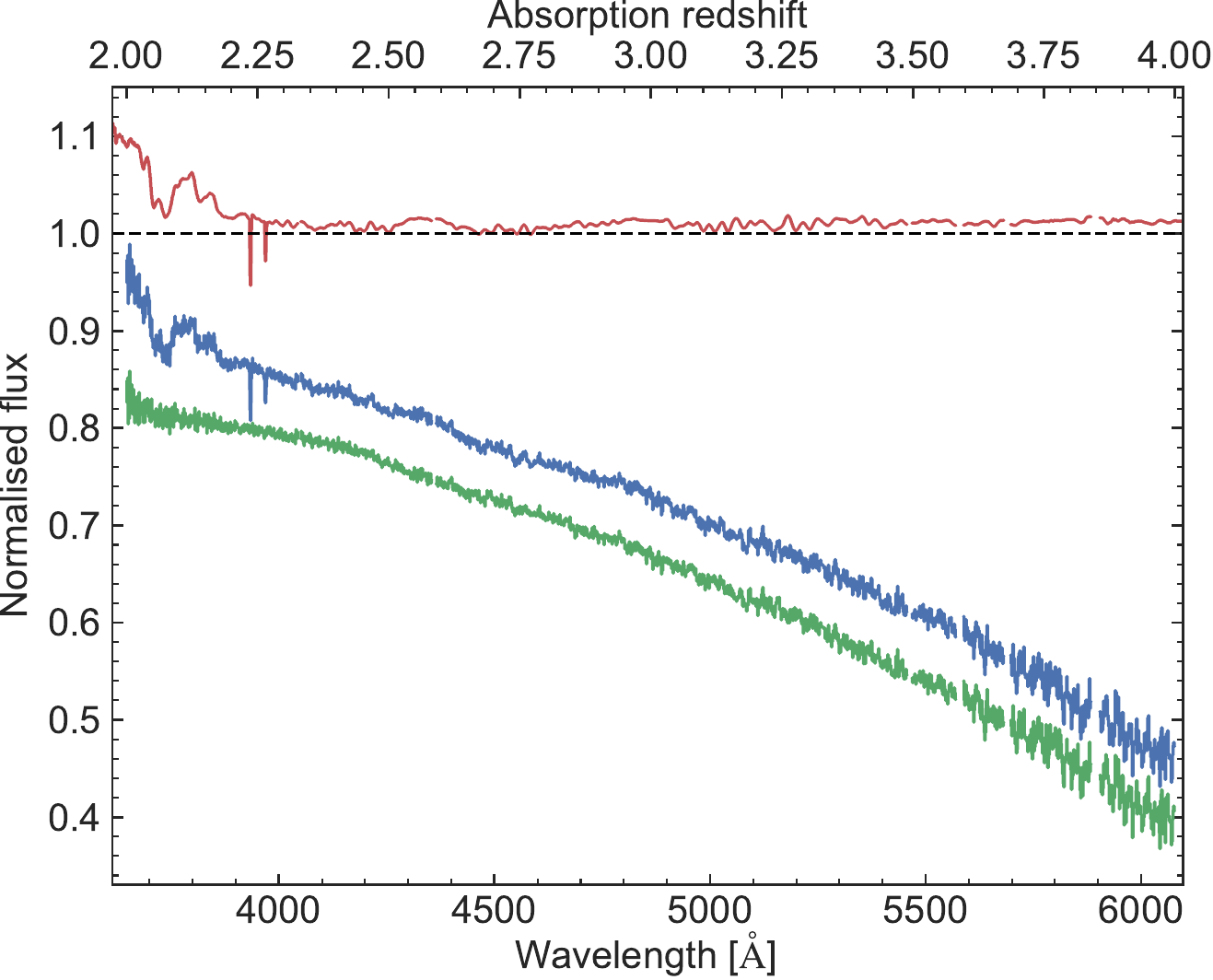}
\caption{The flux calibration correction (red), corrected (green) and uncorrected (blue) mean Ly\,${\alpha}$ forest transmission measured over the entire final sample. A constant vertical offset was given to the blue curve to ease visual inspection. Only the wavelength range corresponding to the absorption redshift range of $2\leq z_{\rm abs}\leq4$ is shown while the correction spans the entire wavelength coverage.}
\label{fig:sdss_flux_calib_corr}
\end{figure}

For each quasar satisfying the four initial selection criteria described in Section~\ref{sec:sample_selection}, we only use pixels with non-zero inverse variance and \texttt{and\_mask} set to 0, uncontaminated by telluric features (the sky mask is used), and outside of the $\pm100$\,{\AA} around each of the emission lines\footnote{The list of emission lines was taken from \url{http://classic.sdss.org/dr7/algorithms/speclinefits.html\#linelist}.}. To avoid division by zero we also exclude pixels where the model is less than 0.1. Pixels with flux deviating from the PCA model by more that $3\sigma$ are discarded. Spectra that have no valid pixels left after all the above filtering is applied or have normalised $\chi^2/{\rm dof} > 1.2$ in the valid pixels are discarded. The correction vector is given by the median of the flux-to-model ratio from the remaining 235,919 quasars. Finally, we smooth the correction by a nine pixel wide boxcar filter. Four sharp features ($3932.0-3938.0$\,{\AA}, $3967.0-3972.5$\,{\AA}, $6320.5-6327.5$\,{\AA} and $9009.0-9035.0$\,{\AA}), including the Galactic Ca\,{\sc ii} absorption, are left unsmoothed and are not used in smoothing the neighbouring pixels. All the quasar spectra in our analysis are divided by this single smoothed correction vector, the smooth red line in Fig.~\ref{fig:sdss_flux_calib_corr}.

We also note a clear turn-down at the Ly\,$\alpha$ forest mean transmission (the way we measure the transmission is described in Section~\ref{sec:mean_lya_transmission}) curve below $z_{\rm abs} \sim 2.5$ shown with the green line in Fig.~\ref{fig:sdss_flux_calib_corr}. The correction function (red line) reveals this effect, which has previously been noted by e.\,g.\ \cite{Bautista2017}. The correction function illustrated is determined directly from the PCA models provided with each observed spectrum. To check this we calculated a new correction function, following a similar procedure as above but instead of using PCA models we used power-law continuum fits and measured the flux to continuum ratio in the continuum fitting regions (continuum fitting is described in Section~\ref{sec:continuum}). The result was very similar to the PCA method.  Both PCA and continuum correction functions reproduce the same small-scale features and also reveal a turn-down around $z_{\rm abs} \sim 2.5$ when applied to the quasar spectra. The reality of this apparent turn-down clearly needs checking further, preferably using independent observations. 

\subsection{Composite spectrum}

\begin{figure*}
\centering
\includegraphics[width=0.75\linewidth]{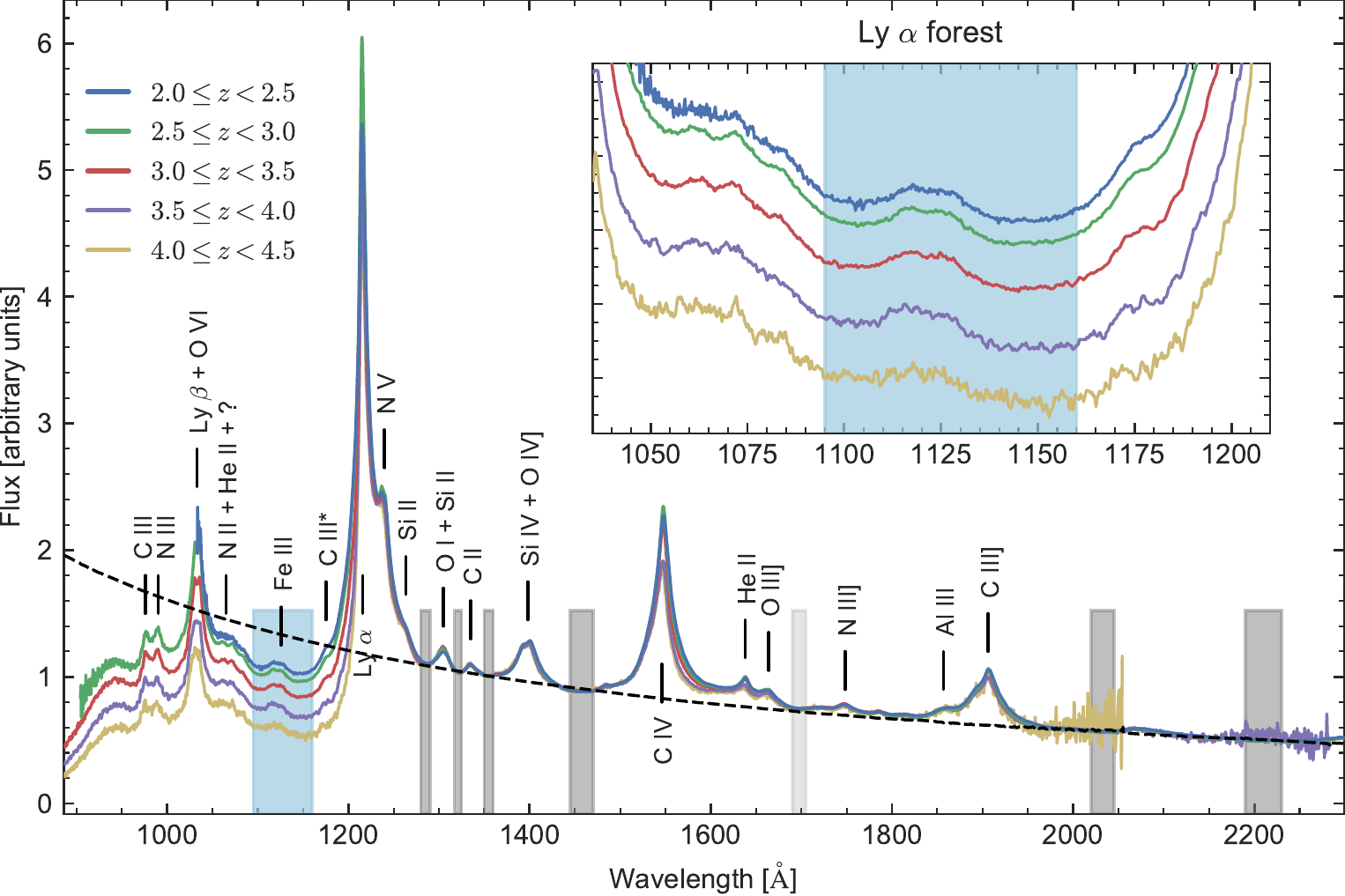}
\caption{Composite quasar spectra for different redshift bins (coloured lines). The black dashed line is the mean power law continuum averaged over all redshift bins. The inset illustrates the wavelength region between Ly\,$\alpha$ and Ly\,$\beta$ emission lines. The Ly\,$\alpha$ forest region where transmission is measured is shown by the shaded blue rectangle. The continuum fitting regions are shown by dark grey rectangles. The light grey region is not used for continuum estimation but the continuum residuals in that region are used as a secondary goodness-of-fit test. The short ticks show the wavelengths of strong emission lines, taken from \protect\cite{VandenBerk2001,Telfer2002}.}
\label{fig:sdss_composite}
\end{figure*}
Fig.~\ref{fig:sdss_composite} shows composite quasar spectra for different redshift bins: $2.0\leq z_{\rm em }<2.5$, $2.5\leq z_{\rm em }<3.0$, $3.0\leq z_{\rm em }<3.5$, $3.5\leq z_{\rm em }<4.0$ and $4.0\leq z_{\rm em }<4.5$. Before co-adding, all the quasars are normalised such that their power-law continua (continuum fitting is described in Section~\ref{sec:continuum}) are equal to 1 at 1360\,{\AA}, shifted to the quasar rest frame and rebinned to the single wavelength scale with the SDSS pixel size of $\approx69$\,km/s. Only quasars that pass the selection criteria (1)-(4) and (6)-(11) defined in Section~\ref{sec:sample_selection} are used. Each spectrum has been divided by the flux calibration correction vector (Section~\ref{sec:corr_vector}) and corrected for dust extinction. A simple (unweighted) mean is used for averaging. Only valid pixels that satisfy the following criteria are used: non-zero inverse variance, zero \texttt{and\_mask}, and not being masked by the sky mask (Section~\ref{sec:sky_mask}). The total number of quasars that contribute to each redshift (from lower to higher) bin is 81{,}597, 47{,}884, 18{,}375, 4265 and 781.

\subsection{Ly\,$\alpha$ forest region used}
\label{sec:lya_forest_region}

The Ly\,$\alpha$ absorption spectral range chosen to measure the H\,{\sc i} characteristics was based on several considerations -- see Fig.~\ref{fig:sdss_composite}. The lower limit of $1095$\,{\AA} avoids the red wings of the Ly\,$\beta$, O\,{\sc vi}, N\,{\sc ii} emission lines and also emission from a line at $\sim1063$\,{\AA} (tentatively ascribed to He\,{\sc ii} plus other unknown lines, \citealt{Telfer2002}). The upper limit $1160$\,{\AA} avoids the blue wings of the C\,{\sc iii}\,$1175$, Ly\,$\alpha$, and N\,{\sc v} emission lines as well as the quasar proximity effect region. The range $1095-1160$\,{\AA} nevertheless contains several weak Fe\,{\sc iii} emission lines. In fact there is essentially no region completely devoid of weak emission. However, since we are interested in comparative measures in this study, that is of no concern.

\subsection{Quasar continuum estimate}
\label{sec:continuum}

To calculate the fraction of light transmitted through the Ly\,$\alpha$ forest, one needs an estimate of the unabsorbed quasar continuum level. This is difficult to estimate locally (i.\,e.\ within the forest itself) due to the spectral resolution and S/N (individual absorption lines are unresolved and blended weak absorption lines mimic a continuum depression). Continuum estimates have previously been derived for SDSS spectra using principal component analysis (PCA) \citep{Lee2012,Lee2013,FontRibera2012,Delubac2015}.  However, the PCA method generally requires a rescaling procedure to normalise the estimated PCA to the local Ly\,$\alpha$ forest flux \citep{Lee2012}.  This rescaling may dilute or even remove any real large-scale anisotropies in the plane of the sky, if present. We therefore instead estimate the Ly\,$\alpha$ forest continuum by fitting a power-law to selected spectral regions longwards of the Ly\,$\alpha$ emission line and extrapolate to the Ly\,$\alpha$ region.

Power-law extrapolation across the Ly\,$\alpha$ emission line also has problems. First, if the intrinsic quasar spectrum is not a single power law but instead is a broken power law \citep{Zheng1997,Telfer2002}, our extrapolated continuum may be placed slightly too high. However, it should be noted that the wavelength at which the putative break occurs is poorly determined \citep{Telfer2002,Binette2008} and could even occur at a lower wavelength than the lowest forest wavelength we use (or may not even exist at all). Moreover, even if such a bias is caused, it can play no role in emulating a cosmological anisotropy. Second, power-law extrapolation does not account for weak emission lines present in the forest, but the same argument concerning emulating cosmological anisotropy applies. Therefore power-law extrapolation is appropriate for our purposes. We fit a power-law $f(\lambda) = C \lambda^{\alpha}$, where $\alpha$ is a spectral index, $C$ is a normalisation constant, and $\lambda$ is wavelength, using selected spectral ranges longwards of the Ly\,$\alpha$ emission line. The following quasar rest-frame regions were used, chosen by inspecting composite spectra -- see Fig.~\ref{fig:sdss_composite}, which appear to be least contaminated by emission lines intrinsic to the quasar itself: $1280-1290$\,{\AA}, $1317-1325$\,{\AA}, $1350-1360$\,{\AA}, $1445-1470$\,{\AA}, $2020-2045$\,{\AA}, $2190-2230$\,{\AA}.

\begin{figure*}
\centering
\includegraphics[width=0.85\linewidth]{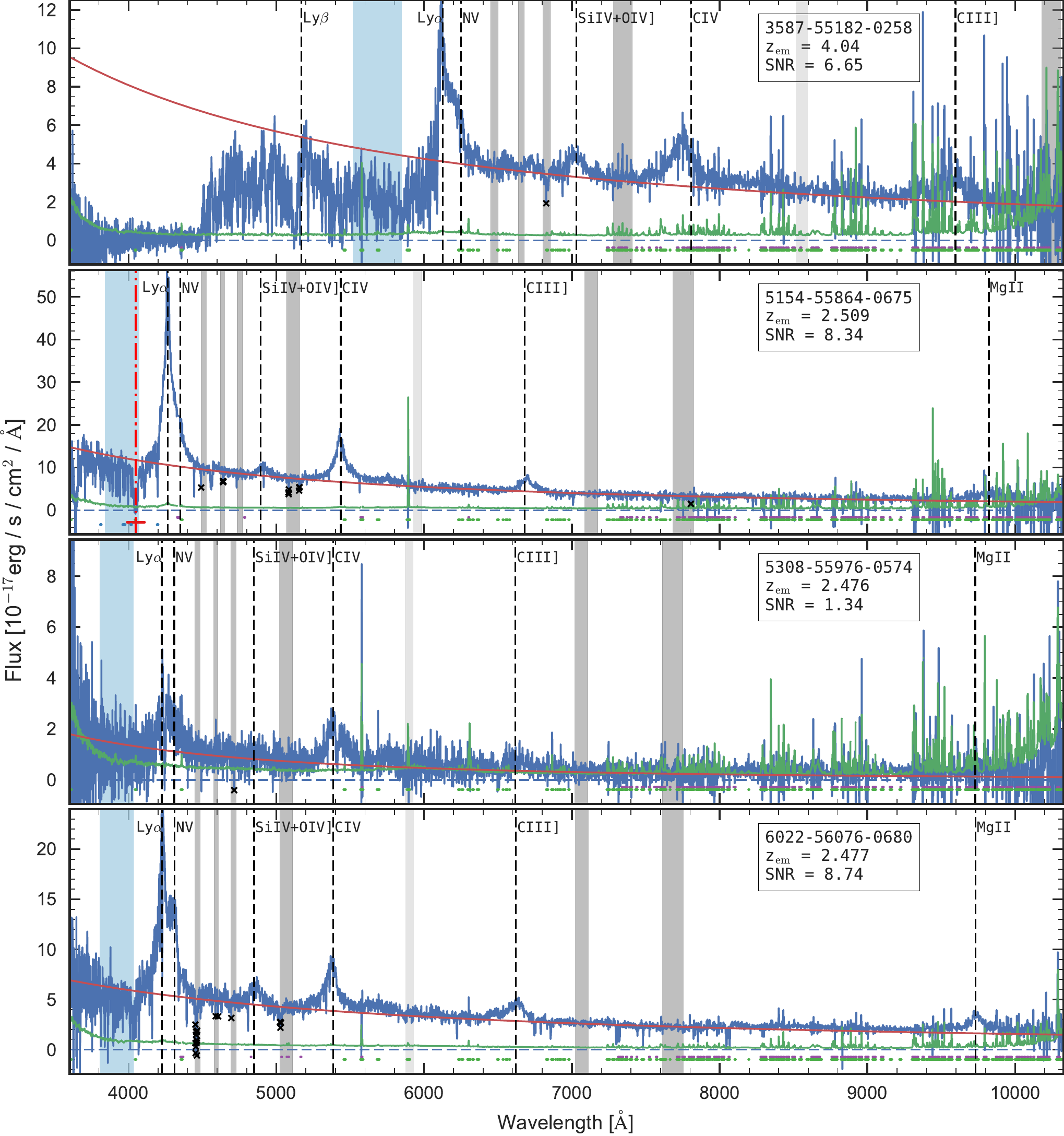}
\caption{Examples of continuum fits for randomly selected quasars from our final sample. In each panel: the blue and green histograms are the flux and the $1\sigma$ error, respectively; the red solid line is the best continuum fit to the flux using the regions indicated by the six dark grey vertical stripes. The extra region used to test goodness of the continuum fit is indicated by the light grey vertical stripe. The black crosses indicate discarded pixels due to metal absorption in the continuum fitting regions. Pixels highlighted with the horizontal lines below $y=0$ are discarded due to the following reasons:  either \texttt{and\_mask}\,$>0$ or zero inverse variance (purple), the sky mask (green), DLA (red, center indicated by a vertical red dash-dotted line) or metal (blue) absorption. The Ly\,$\alpha$ forest region used to measure transmission is shown by the blue vertical band. Strong emission lines are indicated by vertical dashed lines. The quasar plate-MJD-fiber, $z_{\rm em}$ and SNR, are shown in the insets.}
\label{fig:sdss_cont_fit_examples}
\end{figure*}

Unidentified metal absorption lines which happen to fall in any of the continuum fitting segments will bias the continuum fitting placement.  To minimise this, in each of the selected continuum segments, pixels with flux $3\sigma$ below the mean flux in all other pixels in the segment are discarded. This clipping is iterated until convergence, i.\,e.\ until no further pixels are discarded. If, as a result of telluric feature or bad pixel masking or metal clipping, there are less than five valid pixels in a continuum segment, that segment is discarded. Median fluxes and wavelengths in the continuum fitting segments with enough valid pixels are calculated. The final continuum estimate is derived by fitting to these median flux and wavelength points. Four example continuum fits and spectra are illustrated in Fig.~\ref{fig:sdss_cont_fit_examples}.

\subsection{Quasar selection criteria}
\label{sec:sample_selection}

From our core sample of 297,301 quasars, we impose the following selection procedures: (1) Broad Absorption Lines quasars (\texttt{BAL\_FLAG\_VI} set to 1) are rejected; (2) the mean S/N in five SDSS bands (\texttt{SN\_MEDIAN} flag) must be $\geq 1$; (3) quasars should have reliable emission redshifts (\texttt{ZWARNING} set to 0); (4) emission redshifts reported in the quasar catalogues and in the fits files should be within $0.1$ of each other (these first four criteria reduce the sample to 242,348 quasars); (5) given the quasar redshift distribution, the SDSS wavelength coverage and the wavelength ranges of contaminating telluric features, we only use Ly\,$\alpha$ forest data within the absorption redshift range  $2\leq z_{\rm abs} \leq 4$. This constraint reduces the sample further to 145,526 quasars.

\begin{figure}
\centering
\includegraphics[width=1.0\linewidth]{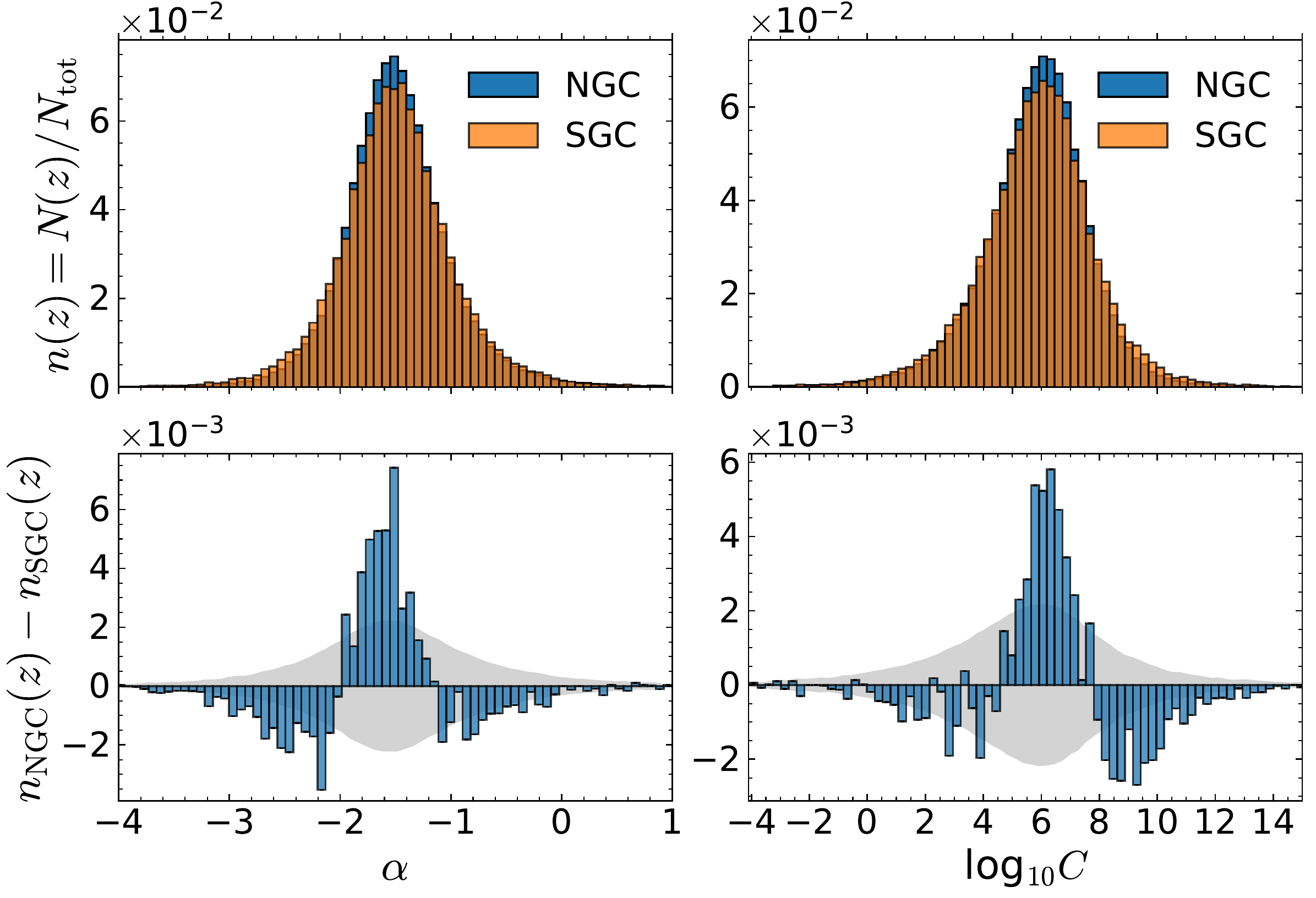}
\caption{Probability distributions of the quasar continuum power law fit parameters ($f(\lambda) = C \lambda^{\alpha}$) for the quasar samples used. {\it Top left:} Spectral index, $\alpha$, NGC (blue) and SGC (orange). The bin width is $\simeq0.073$. 49 quasars with $-6.3<\alpha< -4$ fall outside the range plotted.  {\it Lower left:} The blue histogram illustrates the residuals for the plot above. The grey shaded area illustrates a $\pm 1\sigma$ error estimate.  {\it Top right:} Normalisation constant, $\log_{10} C$, NGC (blue) and SGC (orange). The bin width is $\simeq0.27$. 47 quasars with $15<\log_{10} C<23$ fall outside the range plotted. {\it Lower right:} Analagous to lower left. The $y$-axis on each panel was divided by the value given on top of the panel.}
\label{fig:sdss_si_distribution}
\end{figure}

To remove poor quality spectra, or spectra that may not be quasars, or spectra with problems of various sorts, we imposed six additional criteria aimed at avoiding poor quasar continuum estimates: (6) no more than two continuum fitting regions have less than five valid pixels; (7) there must be a positive median and weighted mean flux in all continuum fitting regions; (8) spectral index calculated using both the median and the weighted mean flux in the continuum fitting regions is less or equal to 1 (to remove extreme values which are likely to be the result of continuum fits that did not work -- see Fig.~\ref{fig:sdss_si_distribution}); (9) the difference between the spectral indices based on median and weighted mean flux in the continuum fitting regions is less or equal to 0.4 (visual inspection of a large number of continuum fits showed that 0.4 was a reasonable choice to strike a balance between ensuring a robust continuum fit and avoiding discarding too many good spectra); (10) the difference between the logarithm of the normalisation constants of the continuum based on median and weighted mean flux in the continuum fitting regions is less or equal to 2 (same justification as above); (11) continuum fitting residuals, based on median and weighted mean flux, in all continuum fitting regions and one extra region ($1690-1705$\,{\AA} in the quasar rest frame) are less or equal to 20 times the spectral error array value (same justification as above).

\begin{figure}
\centering
\includegraphics[width=1.0\linewidth]{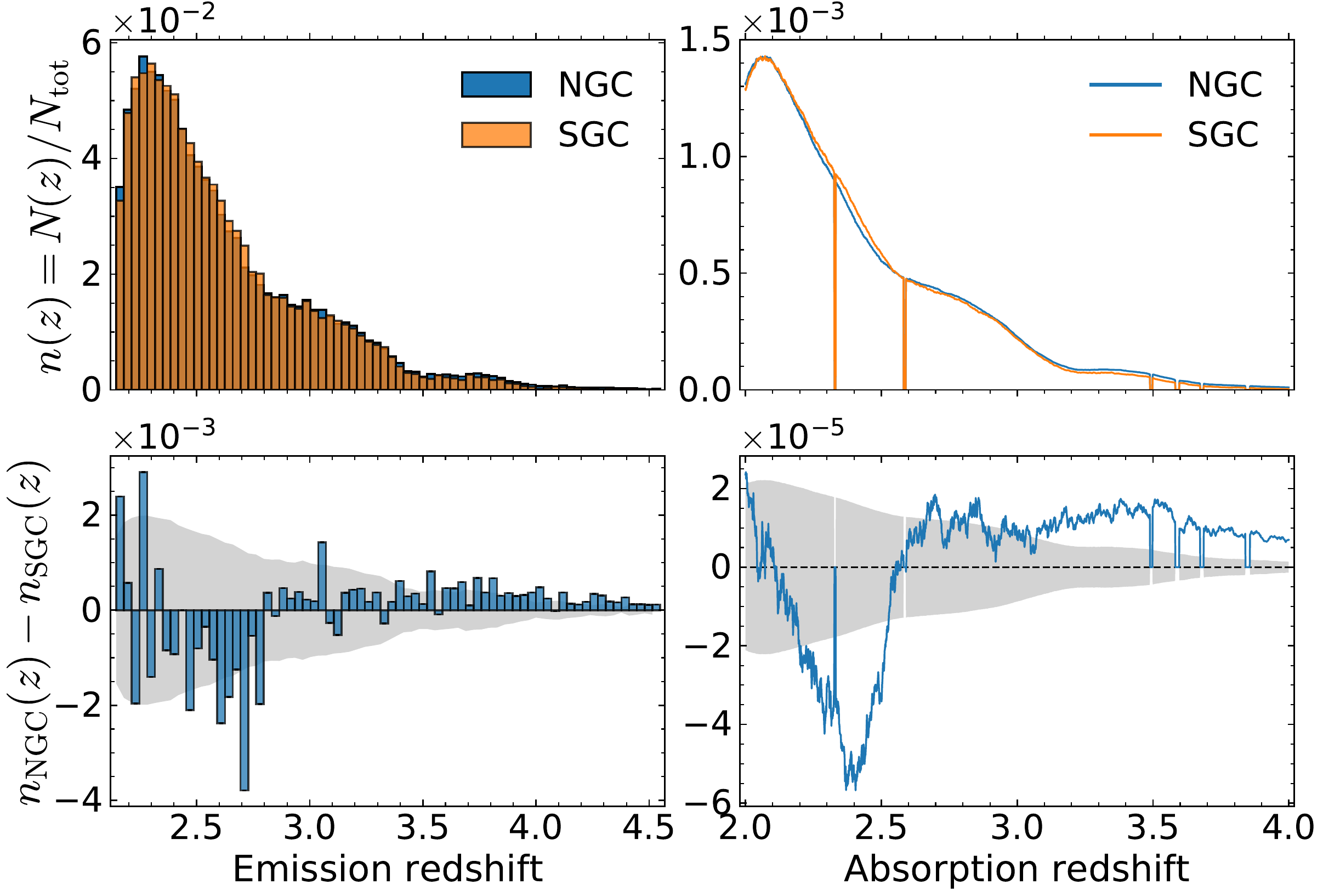}
\caption{{\it Top left:} Emission redshift probability distributions for the quasar samples used, NGC (blue) and SGC (orange). The redshift bin width is $\simeq0.034$. {\it Lower left:} The blue histogram illustrates the residuals for the plot above.  The grey shaded area illustrates a $\pm 1\sigma$ error estimate. {\it Top right:} Number of quasars that contribute to each absorption redshift in the final samples, NGC (blue) and SGC (orange). The sharp drops are pixels masked by the sky mask. {\it Lower right:} Analagous to lower left. The $y$-axis on each panel was divided by the value given on top of the panel.}
\label{fig:sdss_z_distribution}
\end{figure}

The number of quasar spectra that passed all the above selection criteria was 143,652. Of these, 142,661 spectra have valid pixels in their Ly\,$\alpha$ forest region after all the pixel masking steps (due to telluric features, DLAs and pipeline flags) which form our final sample. Probability distributions of the continuum fitting parameters in the final sample, separately for NGC and SGC as well as the difference between the two, are shown in Fig.~\ref{fig:sdss_si_distribution}. Emission and absorption redshift probability distributions are shown in Fig.~\ref{fig:sdss_z_distribution}. Note, neighbouring pixels in the absorption redshift distribution are correlated, the correlation length is defined by the Ly\,$\alpha$ forest region used (Section~\ref{sec:lya_forest_region}).

\section{Analysis}

\subsection{Mean H\,{\sc i} transmission}
\label{sec:mean_lya_transmission}

The simplest H\,{\sc i} absorption measure is the mean absorption, averaged over a large number of contributing pixels.  This was first introduced by \cite{Oke1982}, averaging over the entire Ly\,$\alpha$ forest wavelength range. Here we do not do that but instead maintain the default SDSS spectral pixel size of $\approx 69$\,km/s and compute 
\begin{equation}
\langle F(z) \rangle = \frac{1}{\sum\limits_i \mathpzc{w}_i(z)} \sum_i \mathpzc{w}_i(z) \frac{f^i_{\rm obs}(z)}{f^i_{\rm c}(z)} = \frac{1}{N(z)} \sum_i F_i(z)
\label{eq:F(z)}
\end{equation}
where $F_i(z)=\mathpzc{w}_i(z)f^i_{\rm obs}(z)/f^i_{\rm c}(z)$, $f_{\rm obs}^i(z)$ is the observed flux, and $f^i_{\rm c}(z)$ is the unabsorbed continuum flux, $\mathpzc{w}_i(z)$ is the weight array for the $i^{th}$ quasar where $\mathpzc{w}_i(z)=1$ if the pixel is available and $\mathpzc{w}_i(z)=0$ if not, $N(z)$ is the total number of quasars that contribute to redshift $z$. The summation is taken over all available quasar spectra and is not summed over redshift (as indicated in equation \ref{eq:F(z)} by expressing all quantities as a function of $z$). We have used an unweighted addition (i.e. $\mathpzc{w}_i(z) = 1$ for available pixels) in order to avoid a small number of very high S/N spectra dominating the composite signal, else cosmic variance from individual Ly\,$\alpha$ forest lines (rather than photon counting statistics) remains in the co-added data and hence would be overly important. The Ly\,$\alpha$ absorption spectral range chosen to measure the mean transmission was described in Section~\ref{sec:lya_forest_region}.

\subsection{H\,{\sc i} fluctuations in the plane of the sky}

To explore spatial structure in the plane of the sky, a HEALPix\footnote{\url{http://healpix.sourceforge.net}} visualisation is used. We compute equation~(\ref{eq:F(z)}) for the entire final sample (i.e. NGC + SGC) and then compute the transmission residuals in HEALPix pixel sizes of $0.84$ degrees$^2$ (NSIDE=64). This setting gives a median of 7 quasars per pixel, multiple quasars helping to reduce cosmic variance from individual forest lines at any particular redshift. The transverse distance across one pixel corresponds to a comoving separation in the plane of the sky of $\sim 86$\,Mpc at $z=2$, so spatial information on scales below that is lost.

\begin{figure*}
\centering
\begin{minipage}[h]{0.496\textwidth}
\center{\includegraphics[width=\textwidth]{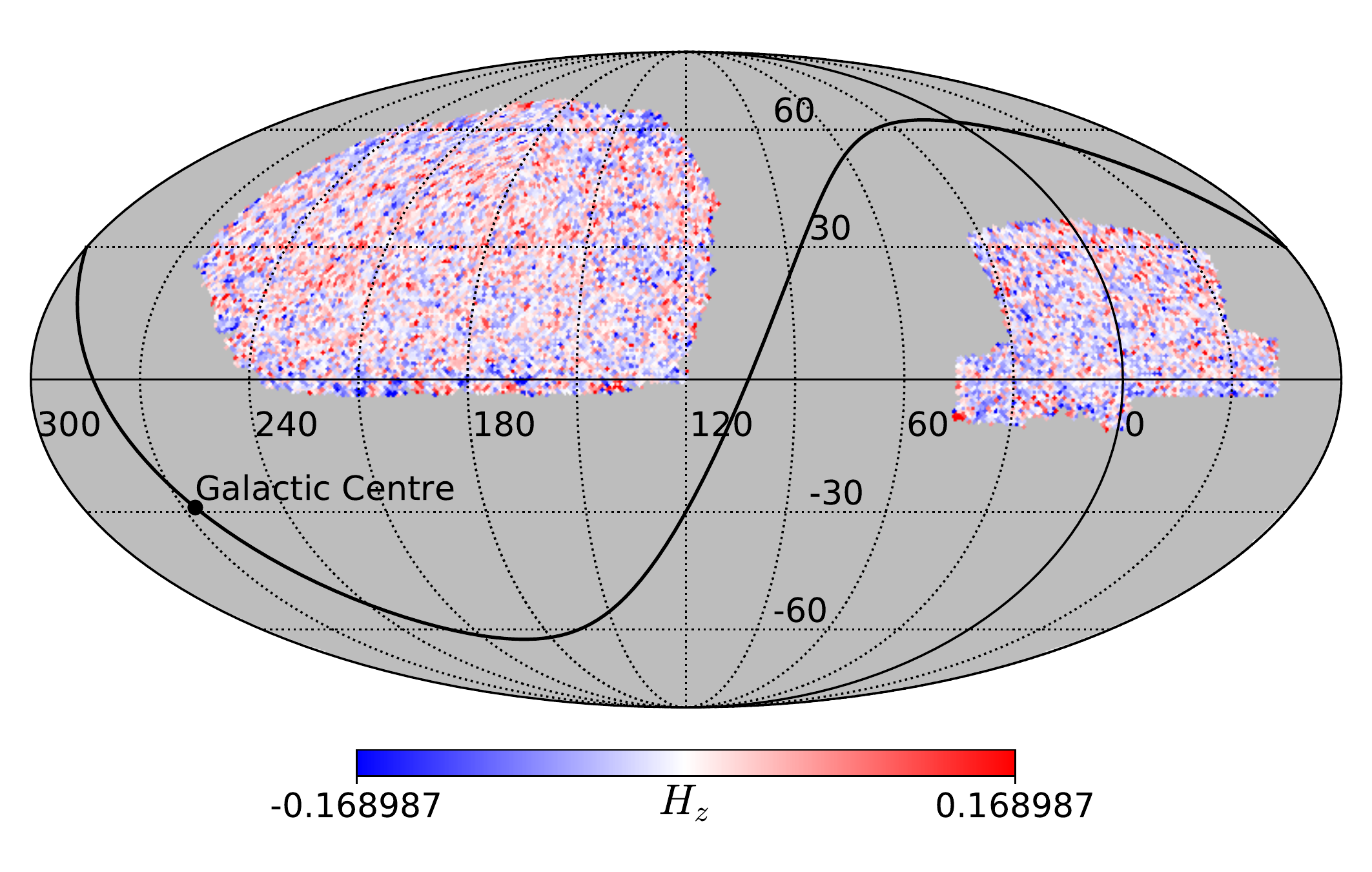}}
\end{minipage}
\hfill
\begin{minipage}[h]{0.496\textwidth}
\center{\includegraphics[width=\textwidth]{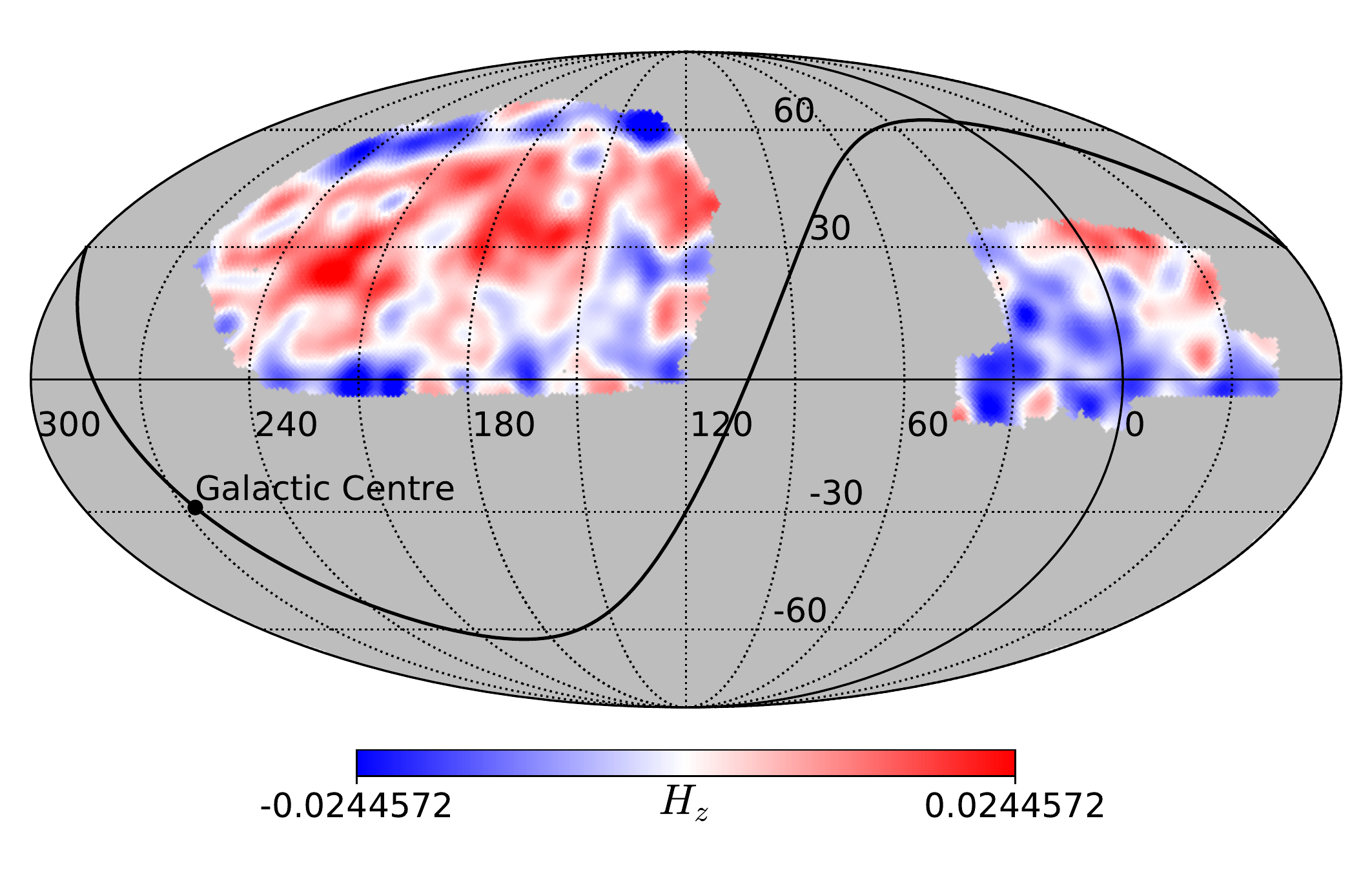}}
\end{minipage}
\vfill
\begin{minipage}[h]{0.496\textwidth}
\center{\includegraphics[width=\textwidth]{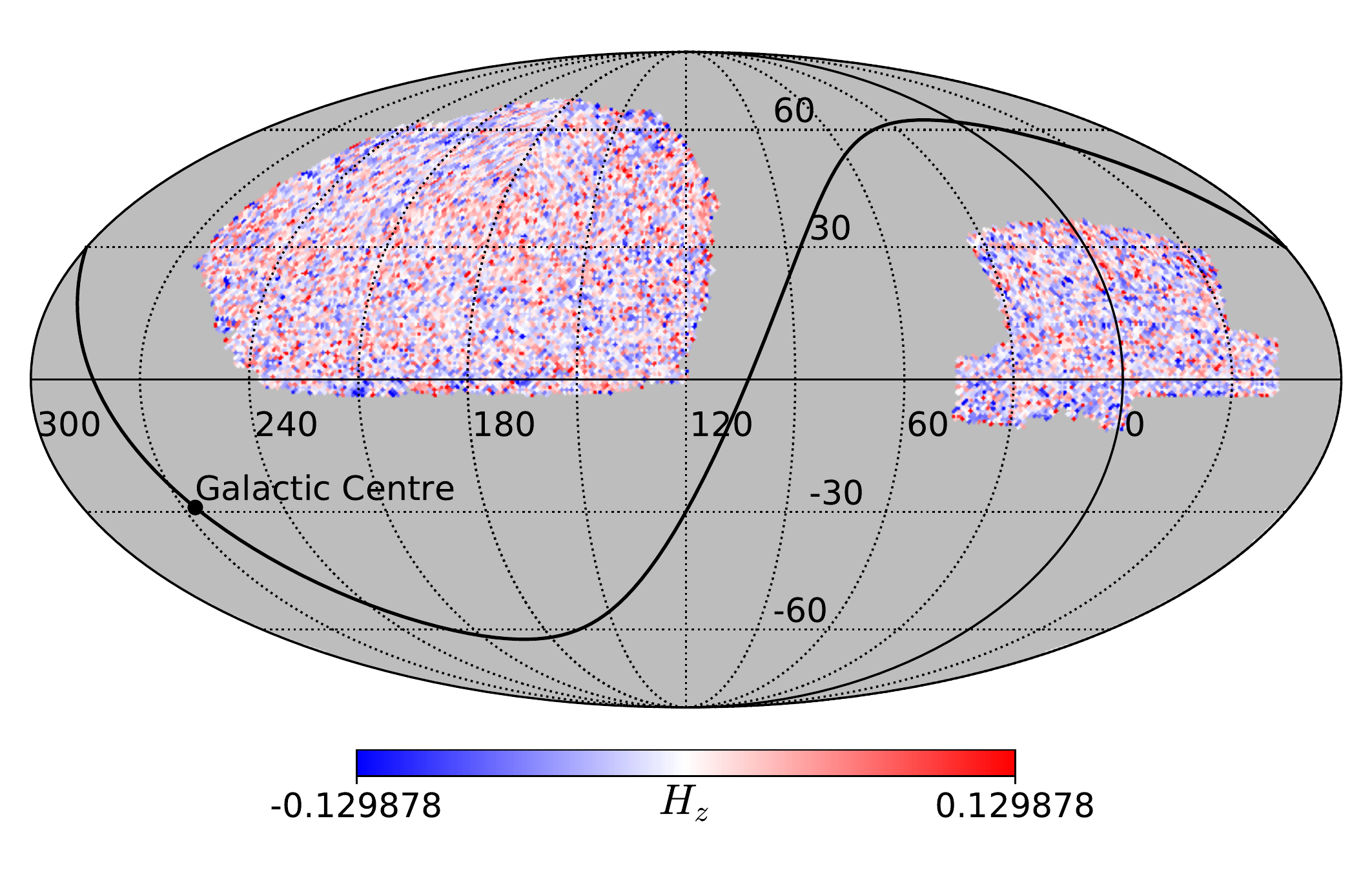}}
\end{minipage}
\hfill
\begin{minipage}[h]{0.496\textwidth}
\center{\includegraphics[width=\textwidth]{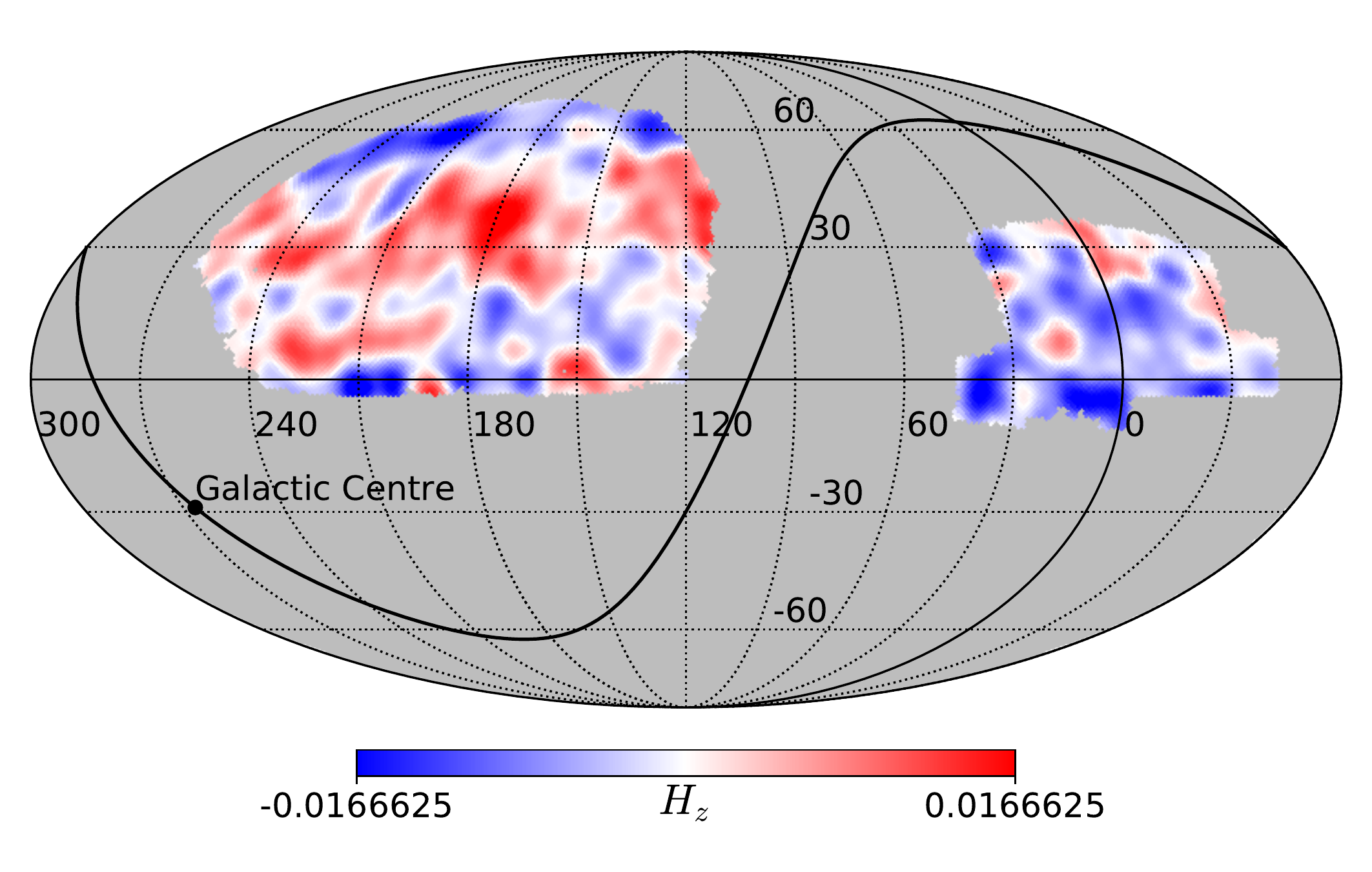}}
\end{minipage}
\caption{HEALPix map with NSIDE=64 of the residual Ly\,$\alpha$ forest mean transmission, $H_z$, calculated over the redshift range of $2\leq z \leq 4$. The equatorial coordinate system is used, with a central point at (120,0) in order to display NGC and SGC without fragmentation. The Galactic plane is shown as a continuous black line passing through the Galactic centre. The dynamic range of each plot is such that 98\% of the data points are within saturation level. The top and bottom panels show maps for the real forest and fake forest datasets, respectively. The right panels are the smoothed versions of the left panels. The smoothing was done with a Gaussian kernel with {\sc fwhm}\,$=7^{\circ}$.}
\label{fig:sdss_healpix_map4}
\end{figure*}

For the $i$th quasar the average transmission residual over the redshift shell $z_1\leq z_{\rm abs}<z_2$ is
\begin{equation}
h_{i,z} = \frac{1}{N_{i,z}}\sum\limits_{z_1}^{z_2} (F_i(z) - \langle F(z) \rangle)
\label{eq:littleh}
\end{equation}
where the summation is taken over all $N_{i,z}$ valid pixels within the redshift shell. If the redshift shell, $\Delta z$, is too small, the residuals are dominated by Ly\,$\alpha$ forest cosmic variance.  The larger the shell-depth $\Delta z$, the more we smooth-out line-of-sight details.  We then average all measurements of $h_{i,z}$ that fall in the same HEALPix pixel on the sky. We label each pixel using a single integer $j$, although $j$ actually maps into the plane of the sky i.e. requires right ascension and declination to define it. For the $j^{th}$ pixel in the HEALPix map we have
\begin{equation}
H_{j,z} = \frac{1}{N_{j,z}}\sum_i^{N_{j,z}} h_{i,j,z}
\label{eq:healpix_map}
\end{equation}
$N_{j,z}$ is the number of quasars belonging to that HEALPix pixel contributing to the required redshift range. HEALPix pixels that have no quasars in them are masked. The set of $\lbrace H_{j,z}\rbrace$ represents a HEALPix map for the redshift shell $z$. Fig.~\ref{fig:sdss_healpix_map4} illustrates the HEALPix map for a summation (in equation \ref{eq:littleh}) over the redshift range of $2<z<4$. HEALPix maps for narrower redshift shells of depth $\Delta z=0.25$ are shown in Appendix~\ref{app:healpix_ps}, Fig.~\ref{fig:sdss_healpix_maps_real} and \ref{fig:sdss_healpix_maps_fake}. The top two panels of Fig.~\ref{fig:sdss_healpix_map4} are derived using the Ly\,$\alpha$ forest regions discussed above. The bottom panels are derived using ``fake forest'' data which we shall discuss shortly. On the left, the data are illustrated in raw HEALPix pixels and on the right a  smoothing using a Gaussian filter with FWHM 7 degrees has been applied. Given the redshift distribution of the SDSS quasars (Fig.~\ref{fig:sdss_z_distribution}), the signal is dominated by the low redshift ($2<z<2.5$) quasars. Red areas indicate greater transmission (less absorption) and blue indicates less transmission (more absorption) compared to the whole-sample mean.

An initial inspection (referring to the top right panel of Fig.~\ref{fig:sdss_healpix_map4}) suggests two characteristics. First, there are low-amplitude correlated regions, extending across angular scales of at least $\sim 40$ degrees, corresponding to $\sim 4000$ Mpc. If real, such structures would be the largest ever detected. Second, the SGC appears visually different to the NGC in that it has slightly more H\,{\sc i} absorption on average. If true, this would indicate cosmological anisotropy.  However, we shall shortly show that both characteristics of the data are unlikely to be real.

\begin{figure*}
\centering
\begin{minipage}[h]{0.496\textwidth}
\center{\includegraphics[width=\textwidth]{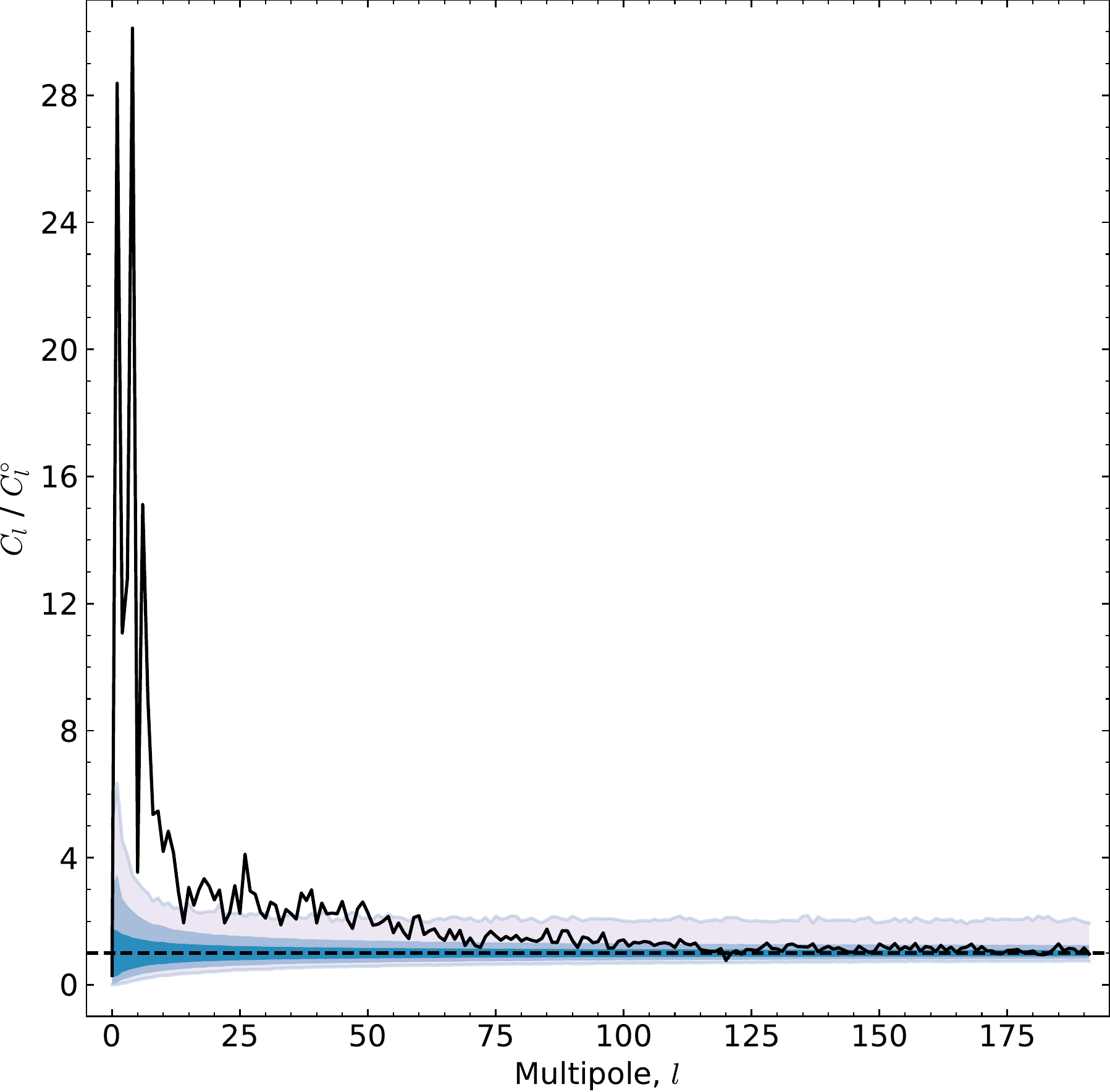}}
\end{minipage}
\hfill
\begin{minipage}[h]{0.496\textwidth}
\center{\includegraphics[width=\textwidth]{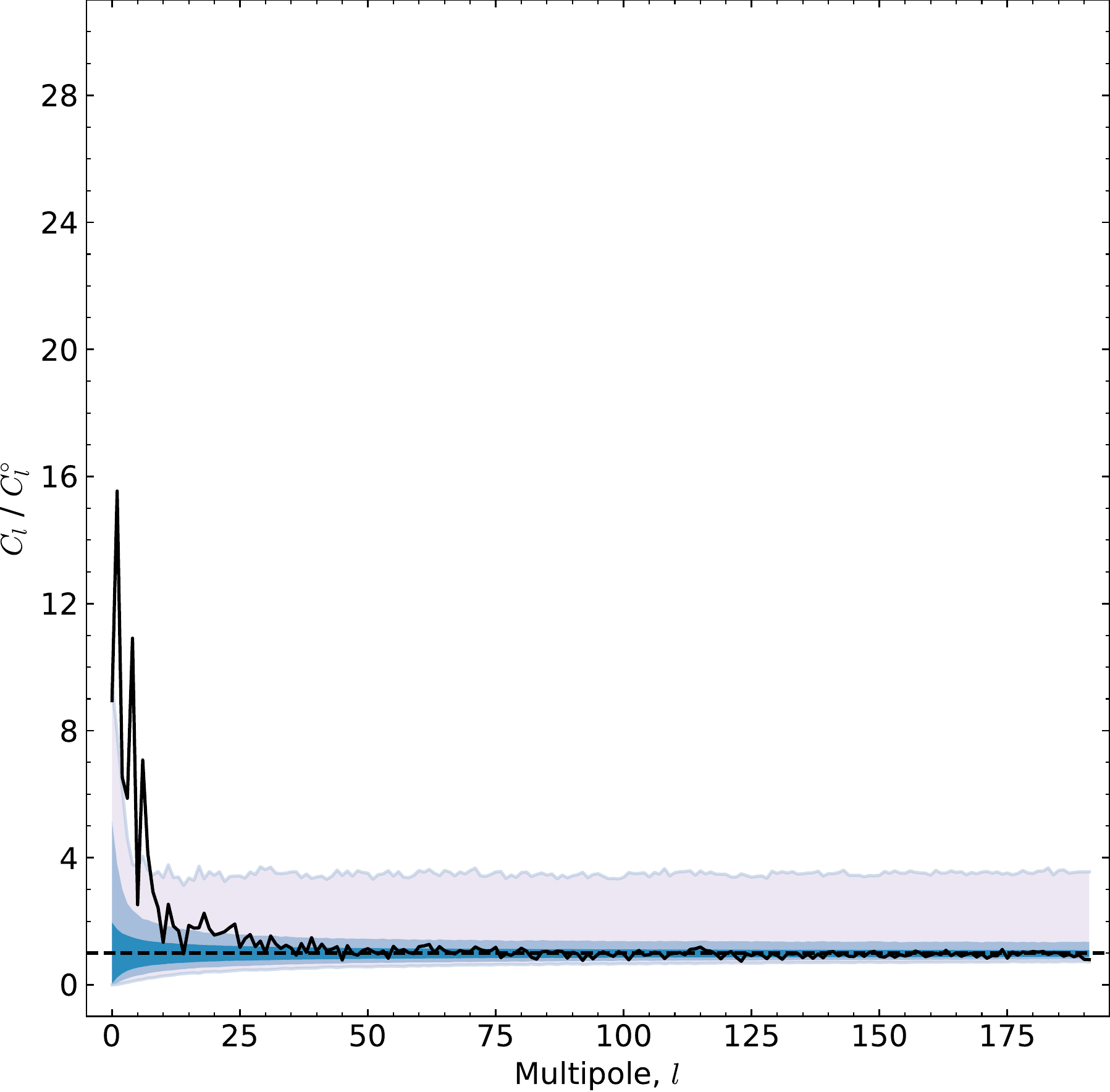}}
\end{minipage}
\caption{The ratio of the power spectrum $C_l$ measured from the HEALPix data to the average power spectrum $C_l^{\circ}$ based on 10{,}000 randomly shuffled samples. The left panel corresponds to the real forest (top left panel in Fig.~\ref{fig:sdss_healpix_map4}) and the right panel corresponds to the fake forest (bottom left panel in Fig.~\ref{fig:sdss_healpix_map4}). The three shaded areas correspond to the $68.3\%$, $95.4\%$ and $99.7\%$ confidence ranges in the randomised samples. Given the MC sample size, the $99.7\%$ contour is very approximate.}
\label{fig:sdss_ps_allz}
\end{figure*}

The fractional transmission fluctuation about the mean value, in the plane of the sky and averaged over a given redshift shell, is
\begin{equation}
\label{eq:rms_healpix}
\left.\frac{\Delta F}{F} \right|_{z} = \frac{1}{\left\langle F_z \right\rangle} \sqrt{\frac{\sum_j H_{j,z}^2}{N_z}}
\end{equation}
where $H_{j,z}$ are the HEALPix map values defined in equation~(\ref{eq:healpix_map}), $N_z$ is the number of valid HEALPix pixels, and $\left\langle F_z \right\rangle$ is the mean transmission for the redshift shell $z$. Computing equation~(\ref{eq:rms_healpix}) for a range of HEALPix pixel sizes finds, as would be expected, that larger pixel sizes produce smaller $\Delta F/F$, since Ly\,$\alpha$ forest cosmic variance becomes small. Degrading the HEALPix resolution of course smoothes the maps and equation~(\ref{eq:rms_healpix}) decreases from $\sim 10$\% to $\sim 3$\% for NSIDE=64 to NSIDE=8.

\subsection{Power spectrum analysis}

The HEALPix \texttt{anafast} function provides power spectra of the HEALPix maps illustrated in Fig.~\ref{fig:sdss_healpix_map4}.  To allow for the SDSS partial sky coverage we Monte Carlo (MC) 10{,}000 maps, each map being a random re-assignment of quasars to the observed sky positions. Fig.~\ref{fig:sdss_ps_allz} illustrates the observed angular power spectrum divided by the mean randomised function for the HEALPix maps shown in Fig.~\ref{fig:sdss_healpix_map4}. The left panel in Fig.~\ref{fig:sdss_ps_allz} is calculated for the map shown in the top left panel of Fig.~\ref{fig:sdss_healpix_map4} (real Ly\,$\alpha$ forest). The right panel is calculated for the map shown in the bottom left panel of Fig.~\ref{fig:sdss_healpix_map4} (fake forest, discussed below). Error contours (shaded areas) are the $68.3\%$, $95.4\%$ and $99.7\%$ confidence ranges in the 10{,}000 randomised samples. The angular power spectra are calculated from ``raw'', i.\,e.\ non-smoothed, maps. Power spectra for HEALPix maps measured in redshift shells of depth $\Delta z=0.25$ are shown in Appendix~\ref{app:healpix_ps} (Figs.~\ref{fig:sdss_ps_real} and \ref{fig:sdss_ps_fake}).

The left hand panel of Fig.~\ref{fig:sdss_ps_allz} suggests the possible presence of significant spatial correlations in the plane of the sky for multipoles $l \lesssim 15$ (or possibly also larger scales), corresponding to angular scales on the sky of greater than $\sim 180/l = 12$ degrees. The power spectrum also suggests possible excess power out to $l \sim 120$. These values seem broadly i.e. visually consistent with the top right HEALPix map in Fig.~\ref{fig:sdss_healpix_map4}.

\subsection{NGC-SGC H\,{\sc i} fluctuations}
\label{sec:ngc_sgc}

Since the SDSS quasar sky coverage is naturally divided into two distinct patches, we compare the mean transmission spectra (equation~\ref{eq:F(z)}) for these two areas. The normalised difference between them is given by
\begin{equation}
\label{eq:DeltaF}
\Delta F_{\textsc{NS}}(z) = \frac{\langle F_{\textsc{ngc}}(z)\rangle - \langle F_{\textsc{sgc}}(z)\rangle}{\sqrt{\sigma_{\textsc{ngc}}^2(z) + \sigma_{\textsc{sgc}}^2(z)}}
\end{equation}
where the $\langle F_{\textsc{ngc}}(z)\rangle$ and $\langle F_{\textsc{sgc}}(z)\rangle$ are given by equation~(\ref{eq:F(z)}), $\sigma_{\textsc{ngc}}(z)$ and $\sigma_{\textsc{sgc}}(z)$ are the standard errors on the mean transmissions at NGC and SGC. The errors were also calculated using a bootstrap simulation ($10{,}000$ trials) and the agreement between the two found to be excellent.

\begin{figure}
\centering
\includegraphics[width=1.0\linewidth]{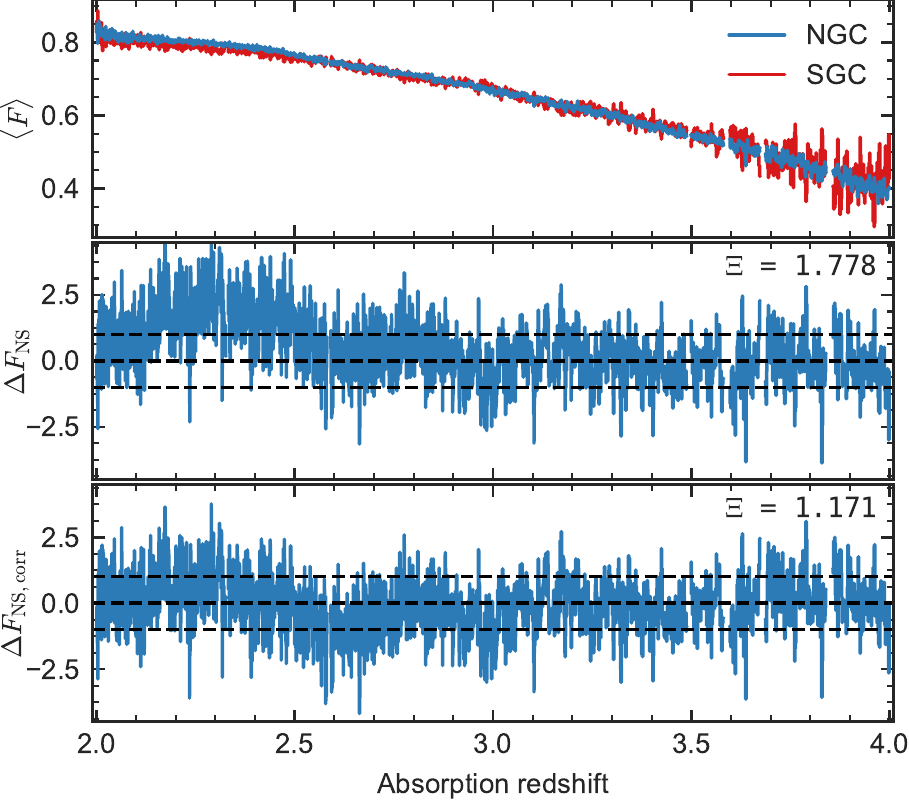}
\caption{{\it Top panel:} Mean transmission measured in the NGC (blue) and SGC (red). {\it Middle panel:} Normalised NGC--SGC mean transmission residuals with $\Xi=1.778$. {\it Bottom panel:} Normalised NGC--SGC mean transmission residuals corrected by the smoothed fake forest residuals with $\Xi=1.171$. The horizontal dashed lines in the middle and bottom panels indicate the $\pm1\sigma$ confidence interval, assuming Gaussian statistics.}
\label{fig:sdss_transm_diff_ngc_sgc}
\end{figure}

The upper panel of Fig.~\ref{fig:sdss_transm_diff_ngc_sgc} illustrates the mean transmissions at NGC and SGC, while the middle panel shows the corresponding normalised residuals (equation~\ref{eq:DeltaF}) between the two. An isotropic universe would be expected to produce residuals which are on average zero.  Consistency between the NGC--SGC mean normalised residuals is parameterised using
\begin{equation}
\Xi = \frac{1}{N}\sum_z\Delta F^2_{\textsc{NS}}(z).
\label{eq:chisq_n_minus_s}
\end{equation}
This quantity is analogous to the usual $\chi^2$ but we have avoided the symbol as the data are not normally distributed. The summation in equation~(\ref{eq:chisq_n_minus_s}) is taken over all valid pixels in the redshift range $2\leq z_{\rm abs}\leq4$ and $N$ is the number of pixels valid in both NGC and SGC in that redshift range. The normalised residuals shown in Fig.~\ref{fig:sdss_transm_diff_ngc_sgc} deviate strongly from zero in the regions $2<z<2.5$. The greater significance at the low redshift end is partially caused by the quasar redshift distributions (Fig.~\ref{fig:sdss_z_distribution}): the $z<2.5$ residuals are far higher S/N than $z>2.5$. To quantify any difference between the $\Xi$ obtained using equation~(\ref{eq:chisq_n_minus_s}) with what we might expect for a perfectly isotropic universe, a MC method is used. We randomly re-assign quasar spectra to the actual quasar sky positions and calculate $\Xi$ as defined in equation~(\ref{eq:chisq_n_minus_s}).  This is repeated 100,000 times to obtain the statistical distribution. As expected, the distribution peaks at unity.  Empirically, in this way, we would conclude that the probability that the observed NGC-SGC normalised residuals could arise from the same parent population is $<10^{-5}$.

The NGC patch is approximately three times larger than the SGC (107{,}920 quasars vs 34{,}741). We thus split the NGC up into three approximately equal-sized patches with $\text{RA}< 163.528$, $163.528\leq\text{RA}<205.528$ and $\text{RA}\geq205.528$. The mean transmission is re-computed for each patch and normalised residuals again compared.  The results of carrying out the same procedure for all combinations of region-pairs suggest the following. First, none of the three NGC patches agree with the SGC patch. The three sets of normalised residuals appear similar to each other in all three cases and similar to the full NGC-SGC set. Second, inter-comparing the three NGC patches with each other, the agreement is far better. An analagous MC test with $\Xi$ summed over the same redshift range ($2<z<4$) does not find such noticeable differences between the three NGC patches. 

The findings above may suggest evidence for anisotropy, although we show shortly that this is unlikely to be so and demonstrate that the data are most likely consistent with an isotropic universe.

\subsection{Support for cosmological isotropy}

Both the SDSS itself and the raw data processing procedures are complex, involving many calibration steps. There are multiple stages at which systematic errors could potentially be introduced into the data. Considerable effort is made to remove or minimise systematics before the data is released to the community for scientific analysis \citep{Bolton2012, Abolfathi2018} but it is nevertheless possible some remain. If systematics are present in the data that could be responsible for the putative Ly\,$\alpha$ anisotropy, this ought to be echoed in other rest-frame sections of the quasar spectra. To explore this we defined a ``fake forest'' region close in wavelength to the real Ly\,$\alpha$ forest but at wavelengths just longwards of the Ly\,$\alpha$ emission line. We previously selected six continuum fitting regions longwards of the Ly\,$\alpha$ emission line (Fig.~\ref{fig:sdss_composite}). We retain the three rightmost regions as continuum fitting regions but now use the two leftmost regions as ``fake forest'' regions, analysing those in the same way as the real forest data. We discard one region so as to closely match the continuum extrapolation distances for fake and real samples. We note some differences between the real and fake forest measurements. Firstly we use only three continuum fitting regions for the fake forest but six for the real forest. The continuum should thus be better determined for the real forest. Secondly, the fake forest pixels are at higher rest-frame wavelength, so the initial quasar selection is slightly different and results in a higher redshift sample overall. Apart from these differences, the fake forest test should emulate the joint impact of all systematics in the data.

The HEALPix map for the fake forest dataset is shown in the lower panels of Fig.~\ref{fig:sdss_healpix_map4}. Both the real and fake datasets reveal similar anisotropic signals, suggesting the putative anisotropy is caused by a systematic in the data. The power spectrum for the fake forest dataset is shown in the right panel of Fig.~\ref{fig:sdss_ps_allz}. The left (real forest) and right (fake forest) power spectra in this Figure appear similar. There are nevertheless two interesting differences: (i) the power spectrum amplitude for the real forest is approximately twice as high at low multipoles ($\lesssim 15$) as the fake forest, and (ii) the real forest power spectrum suggests non-zero power out to around $l \sim 120$ whilst the fake forest does not. However, given the MC error contours, these apparent differences may suggest that despite the spatial systematics present in the data, there may be some residual signal indicating large-scale Ly\,$\alpha$ clustering. However, this cannot be considered compelling and may simply be spurious or a consequence of the differences between the real and fake forest data discussed above.

We test quantitatively whether the non-zero normalised residuals seen in the real forest NGC-SGC transmission curves can be explained by systematics revealed in the fake data as follows. First, the fake forest non-normalised residuals are smoothed using a 501 pixel wide boxcar filter
\begin{equation}
\label{eq:fake_corr}
\phi(z) = \textrm{boxcar}_{501}\left(\langle F^{\textsc{fake}}_{\textsc{ngc}}(z)\rangle - \langle F^{\textsc{fake}}_{\textsc{sgc}}(z)\rangle\right)
\end{equation}
where the $F$'s are measured using the fake forest data. The real (non-normalised) NGC-SGC residual curve is corrected by subtracting $\phi(z)$. The corresponding corrected normalised residuals are shown in the bottom panel of Fig.~\ref{fig:sdss_transm_diff_ngc_sgc}, calculated using equation~(\ref{eq:DeltaF}). The important point to note here is the drop in the value of $\Xi$ (equation~\ref{eq:chisq_n_minus_s}) from 1.778 to 1.171.  Correcting the real forest residuals by the fake forest curve reveals the presence of a position-dependent systematic error in the data that can emulate large-scale structure.

Given this fake forest correction, we MC the above process by randomly shuffling the real and the fake datasets 100,000 times. Each time, a new correction (equation~\ref{eq:fake_corr}) is measured and is subtracted from the real forest (randomised). We find that $8\%$ of the random realisations have $\Xi$ greater or equal to the value measured in the unshuffled data.  The interpretation of this is that the real forest NGC and SGC transmission curves are consistent with each other and hence that the data are consistent with an isotropic universe.

Comparing the observed H\,{\sc i} transmission curves for the two volumes delineated by the redshift range covered and the NGC and SGC sky areas provides a volume-specific measure of isotropy as follows. The fractional difference between the mean NGC and SGC transmission curves is
\begin{equation}
\label{eq:wm_transm_diff}
X_{\textsc{NS}} = \frac{1}{\sum\limits_{z} w(z)} \sum\limits_{z} w(z) \frac{\langle F_{\textsc{ngc}}(z)\rangle - \langle F_{\textsc{sgc}}(z)\rangle - \phi(z)}{\textrm{boxcar}_{51}(\langle F(z)\rangle)}.
\end{equation}
The denominator on the right hand side is the mean transmission over the whole sample (NGC\,$+$\,SGC), smoothed using a 51 pixel wide boxcar filter. The weights above are $w(z) = \textrm{boxcar}_{51}(\langle F(z)\rangle)^2/(\sigma^2_{\textsc{ngc}}(z) + \sigma^2_{\textsc{sgc}}(z))$ where the $\sigma$'s are the same as in equation~(\ref{eq:DeltaF}).

After the real forest transmission curves have been corrected using the fake forest transmission curve, we get $X_{\textsc{NS}} = -7.6 \times 10^{-4}$. To quantify the significance of this value we ran a MC test with 100,000 realisations, randomly re-allocating quasar spectra to existing sky positions. This results in $71\%$ of the MC realisations with values of $X_{\textsc{NS}}$ as large or larger than the observed value (the comparison is done using absolute values) i.\,e.\ the two volumes subtended by NCG and SGC are consistent with each other, implying that there is no evidence for any departure from an isotropic universe.

\section{Conclusions}

We have used the largest available sample of Ly\,$\alpha$ forest data to check on cosmological isotropy. The cosmological scales explored are the largest to date over the redshift range $2<z<4$. Whilst the raw data indicate anisotropy, a detailed assessment of potential systematics shows that the data are consistent with an isotropic universe and hence with the standard $\Lambda$CDM cosmological model. This conclusion is independent of results obtained using cosmic microwave background data.  

A careful revision of the flux calibration procedures in the SDSS data reduction pipeline is needed to remove the sky position dependent systematics we have discovered. Any further studies that attempt to measure the three dimensional structure of neutral hydrogen over cosmological scales using SDSS Ly\,$\alpha$ forest data need to account for spatially correlated systematics. 

Natural follow-on studies from the work presented in this paper are cross-correlation tests with other matter distributions such as the CMB and quasar distributions. We have not attempted this primarily because, given the spatial systematics found, these systematics would merely be propagated into cross-correlation analyses and the results would be difficult to interpret.

Future surveys will impose tighter limits provided observational strategies are carefully considered. Such surveys include LSST\footnote{\url{https://www.lsst.org/}}, 4MOST \citep{Jong2012}, DESI \citep{DESI2016}, MSE \citep{Percival2019}, WEAVE-QSO \citep{Pieri2016}. We have seen in this analysis that experimental inaccuracies (perhaps associated with astrometry errors or fibre-positioning errors) can lead to quasar spectral shape systematics that vary with sky position. The design of new surveys should take particular care not to repeat this.

\section*{Acknowledgements}

We thank an anonymous referee for useful comments that helped to improve the submitted text. We have enjoyed interesting and fruitful discussions with Signe Riemer-S{\o}rensen, John Barrow, Dinko Milakovic, Chung-Chi Lee, Joe Wolfe, Julian Bautista, Andreu Font-Ribera, Alec Boksenberg, and Lluis Mas-Ribas whilst preparing this manuscript. This work also benefited from a previous (unpublished) student project with Jon Ouellet. EOZ is supported by an Australian Government Research Training Program (RTP) Scholarship and is grateful to GitHub for the Student Developer Pack. JKW thanks the John Templeton Foundation for support, the Department of Applied Mathematics and Theoretical Physics and the Institute of Astronomy Cambridge for hospitality and support, and Clare Hall Cambridge for a Visiting Fellowship. This research made use of the Python packages: astropy \citep{Astropy2013}, dustmaps \citep{Green2018}, extinction \citep{Barbary2016}, healpy \citep[HEALPix;][]{Gorski2005,Zonca2019}, h5py\footnote{\url{https://www.h5py.org/}}, matplotlib \citep{Hunter2007}, numpy \citep{VanDerWalt2011}, pandas \citep{McKinney2010}, scipy \citep{Oliphant2007} and NASA's Astrophysics Data System.

SDSS-III\footnote{\url{http://www.sdss3.org/}} funding is from the Alfred P. Sloan Foundation, the Participating Institutions, the National Science Foundation, and the U.S. Department of Energy Office of Science. SDSS-III is managed by the Astrophysical Research Consortium for the Participating Institutions of the SDSS-III Collaboration including the University of Arizona, the Brazilian Participation Group, Brookhaven National Laboratory, Carnegie Mellon University, University of Florida, the French Participation Group, the German Participation Group, Harvard University, the Instituto de Astrofisica de Canarias, the Michigan State/Notre Dame/JINA Participation Group, Johns Hopkins University, Lawrence Berkeley National Laboratory, Max Planck Institute for Astrophysics, Max Planck Institute for Extraterrestrial Physics, New Mexico State University, New York University, Ohio State University, Pennsylvania State University, University of Portsmouth, Princeton University, the Spanish Participation Group, University of Tokyo, University of Utah, Vanderbilt University, University of Virginia, University of Washington, and Yale University.




\bibliographystyle{mnras}
\bibliography{mybibliography} 



\appendix

\section{Spatial structures in different redshift shells}
\label{app:healpix_ps}

Here we present detailed versions of Fig.~\ref{fig:sdss_healpix_map4} and \ref{fig:sdss_ps_allz} for different redshift shells of depth $\Delta z=0.25$. 

\begin{figure*}
\centering
\begin{minipage}[h]{0.495\textwidth}
\center{\includegraphics[page=1, width=\textwidth]{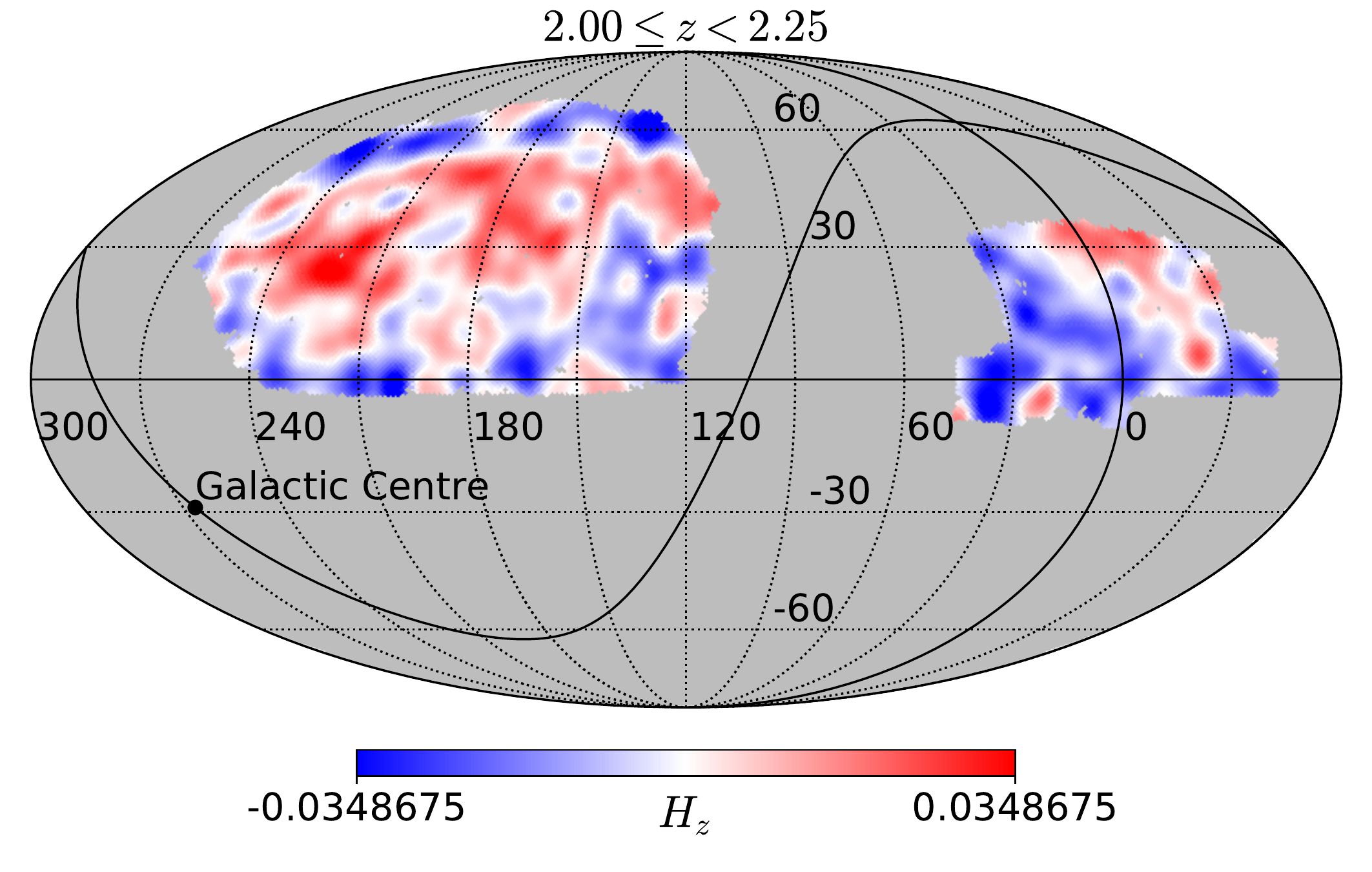}}
\end{minipage}
\hfill
\begin{minipage}[h]{0.495\textwidth}
\center{\includegraphics[page=2, width=\textwidth]{figures/healpix6.pdf}}
\end{minipage}
\vfill
\begin{minipage}[h]{0.495\textwidth}
\center{\includegraphics[page=3, width=\textwidth]{figures/healpix6.pdf}}
\end{minipage}
\hfill
\begin{minipage}[h]{0.495\textwidth}
\center{\includegraphics[page=4, width=\textwidth]{figures/healpix6.pdf}}
\end{minipage}
\vfill
\begin{minipage}[h]{0.495\textwidth}
\center{\includegraphics[page=5, width=\textwidth]{figures/healpix6.pdf}}
\end{minipage}
\hfill
\begin{minipage}[h]{0.495\textwidth}
\center{\includegraphics[page=6, width=\textwidth]{figures/healpix6.pdf}}
\end{minipage}
\caption{HEALPix maps with NSIDE=64 of the residual Ly\,$\alpha$ (real) forest mean transmission, $H_z$, calculated over different redshift shells of depth $\Delta z=0.25$. The dynamic range of each plot is such that 98\% of the data points are within saturation level. The smoothing was done with a Gaussian kernel with {\sc fwhm}\,$=7^{\circ}$.}
\label{fig:sdss_healpix_maps_real}
\end{figure*}

\begin{figure*}
\centering
\includegraphics[width=\textwidth]{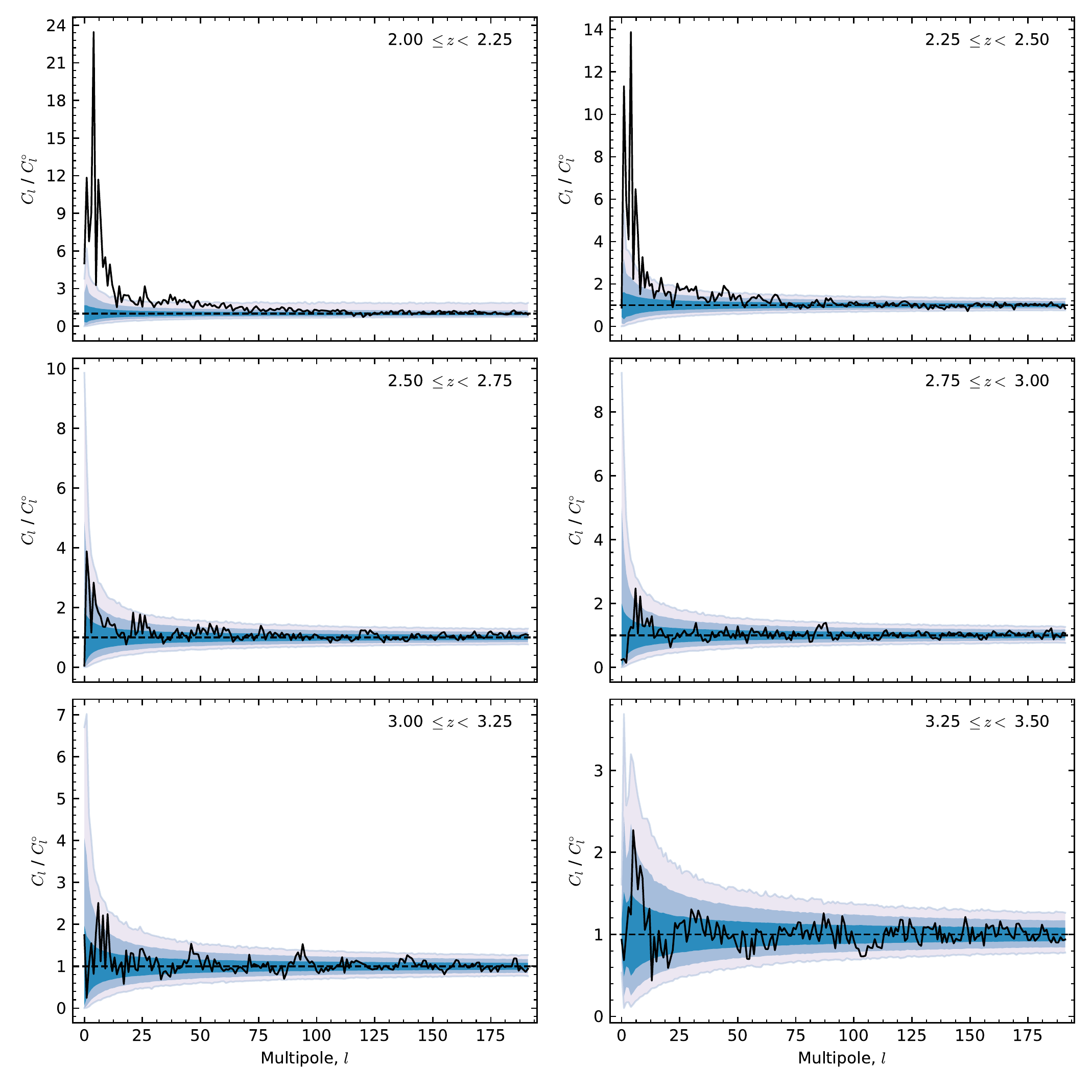}
\caption{The normalised power spectra for the (unsmoothed) HEALPix maps shown in Fig.~\ref{fig:sdss_healpix_maps_real}. The shaded areas indicating error contours are as described in Fig.~\ref{fig:sdss_ps_allz}}
\label{fig:sdss_ps_real}
\end{figure*}

For the real forest, the HEALPix maps and the corresponding normalised power spectra are shown in Fig.~\ref{fig:sdss_healpix_maps_real} and \ref{fig:sdss_ps_real}, respectively. The high redshift shells ($z>3.5$) are not shown due to the rapid fall of the number of quasars at these redshifts. For redshift shells $3<z<3.25$ and $3.25<z<3.5$, since its HEALPix maps are sparse and noisy, we discard $2\%$ of outliers (spectra with equation~\ref{eq:littleh} outside the $98\%$ quantiles) when calculating the power spectra given in the bottom panels of Fig.~\ref{fig:sdss_ps_real}.

\begin{figure*}
\centering
\begin{minipage}[h]{0.495\textwidth}
\center{\includegraphics[page=1, width=\textwidth]{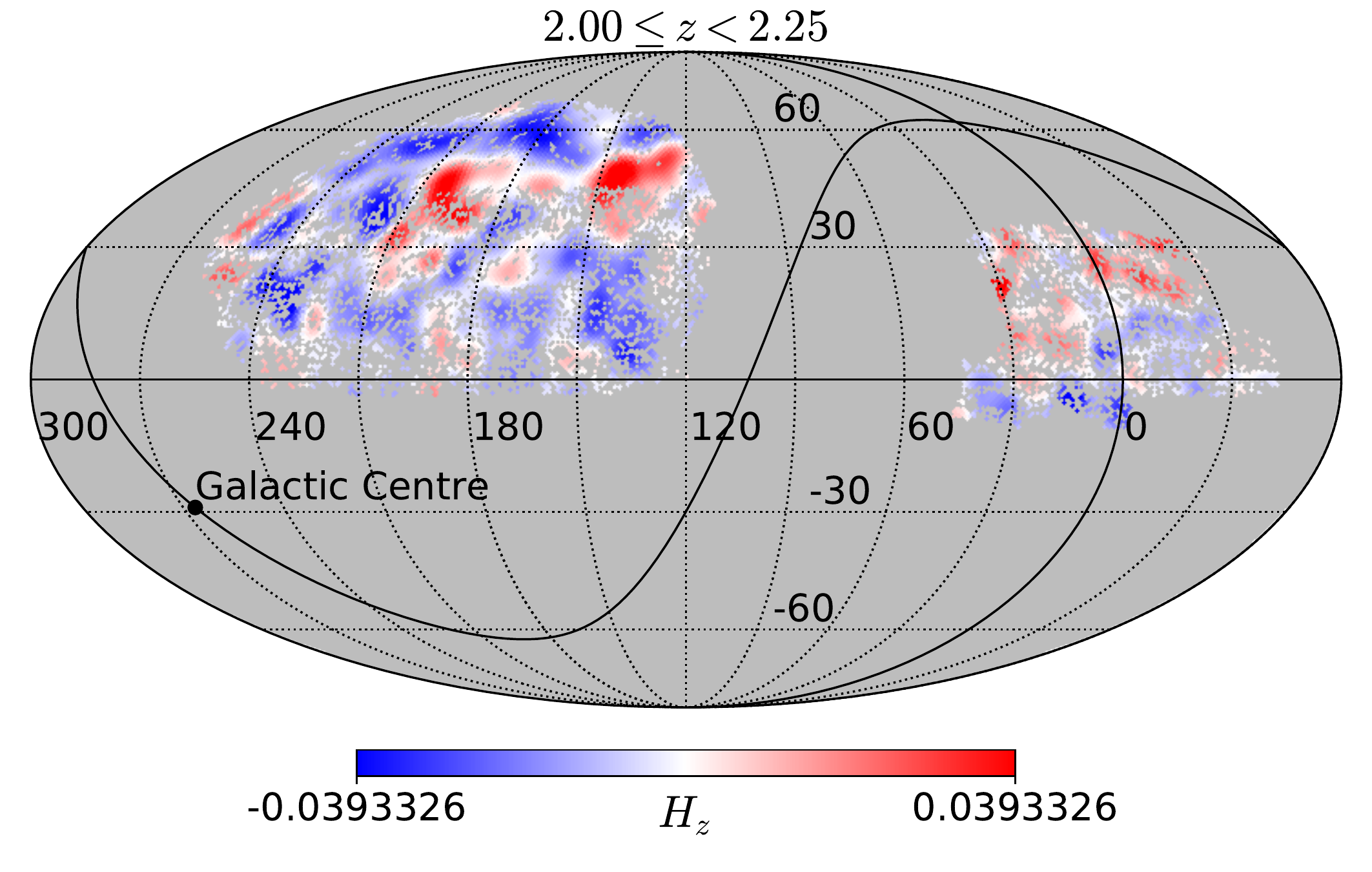}}
\end{minipage}
\hfill
\begin{minipage}[h]{0.495\textwidth}
\center{\includegraphics[page=2, width=\textwidth]{figures/fake_healpix_contreg2.pdf}}
\end{minipage}
\vfill
\begin{minipage}[h]{0.495\textwidth}
\center{\includegraphics[page=3, width=\textwidth]{figures/fake_healpix_contreg2.pdf}}
\end{minipage}
\hfill
\begin{minipage}[h]{0.495\textwidth}
\center{\includegraphics[page=4, width=\textwidth]{figures/fake_healpix_contreg2.pdf}}
\end{minipage}
\vfill
\begin{minipage}[h]{0.495\textwidth}
\center{\includegraphics[page=5, width=\textwidth]{figures/fake_healpix_contreg2.pdf}}
\end{minipage}
\hfill
\begin{minipage}[h]{0.495\textwidth}
\center{\includegraphics[page=6, width=\textwidth]{figures/fake_healpix_contreg2.pdf}}
\end{minipage}
\caption{
HEALPix maps with NSIDE=64 of the residual fake forest mean transmission, $H_z$, calculated over different redshift shells of depth $\Delta z=0.25$. The dynamic range of each plot is such that 98\% of the data points are within saturation level. The smoothing was done with a Gaussian kernel with {\sc fwhm}\,$=7^{\circ}$.}
\label{fig:sdss_healpix_maps_fake}
\end{figure*}

\begin{figure*}
\centering
\includegraphics[page=1, width=\textwidth]{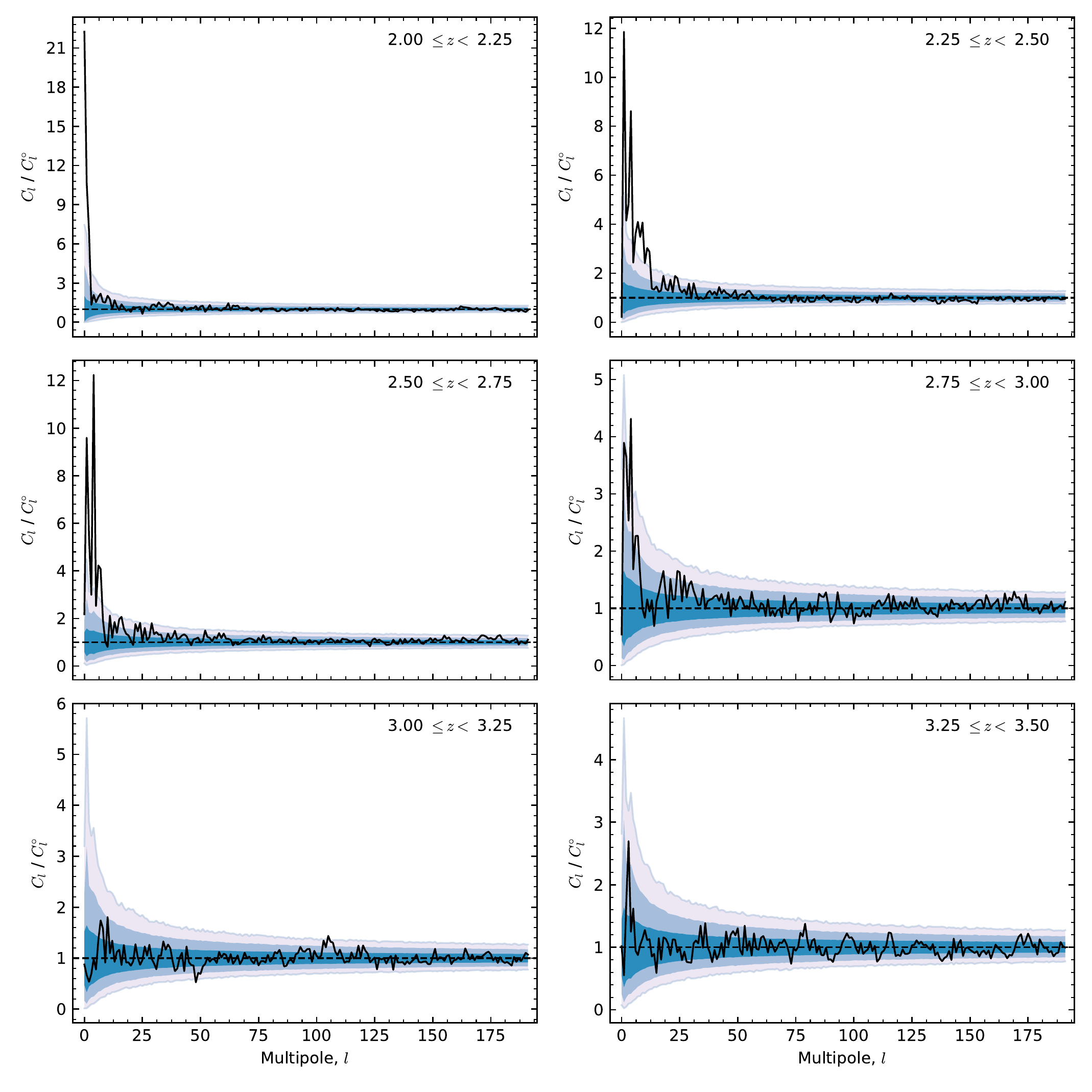}
\caption{The normalised power spectra for the (unsmoothed) HEALPix maps shown in Fig.~\ref{fig:sdss_healpix_maps_fake}. The shaded areas indicating error contours are as described in Fig.~\ref{fig:sdss_ps_allz}}
\label{fig:sdss_ps_fake}
\end{figure*}

For the fake forest, the HEALPix maps and the corresponding normalised power spectra are shown in Fig.~\ref{fig:sdss_healpix_maps_fake} and \ref{fig:sdss_ps_fake}, respectively. The absorption redshift distribution is different for the real and fake forests because of the way in which the latter is necessarily defined. This means that corresponding redshift shell HEALPix maps and power spectra for corresponding redshift shells have different S/N ratios. For the lowest redshift shell ($2<z<2.25$), since its HEALPix map is sparse and noisy, we discard $2\%$ of outliers (spectra with equation~\ref{eq:littleh} outside the $98\%$ quantiles) when calculating the power spectrum given in the top left panel of Fig.~\ref{fig:sdss_ps_fake}.

\bsp	
\label{lastpage}
\end{document}